\documentclass[aos,MSNbibl,nameyear,dvips]{arximspdf}
\usepackage{dcolumn}
\usepackage{graphicx}

\doi{10.1214/14-AOS1220} 
\volume{42}
\issue{5}
\pubyear{2014}
\firstpage{1693}
\lastpage{1724}
\docsubty{FLA}

\makeatletter
\newcolumntype{d}[1]{D{.}{.}{#1}}

\newcommand{\rrvert}{\vert}
\newcommand{\rrVert}{\Vert}
\newcommand{\llvert}{\vert}
\newcommand{\llVert}{\Vert}
\newcommand{\Rhat}{\widehat R}
\newcommand{\Qhat}{\widehat Q}
\newcommand{\gv}{|}
\newcommand{\Var}{\operatorname{Var}}
\newcommand{\logit}{\operatorname{logit}}
\newcommand{\diag}{\operatorname{diag}}
\newtheorem{theorem}{Theorem}
\newtheorem{corollary}[theorem]{Corollary}
\newtheorem{lemma}[theorem]{Lemma}
\newtheorem{proposition}[theorem]{Proposition}
\newproclaim{ex}{Example}
\makeatother

\begin{document}
\begin{frontmatter}

\title{Local case-control sampling: Efficient subsampling in
imbalanced data sets}
\runtitle{Local case-control sampling}

\begin{aug}
\author{\fnms{William}~\snm{Fithian}\corref{}\ead[label=e1]{wfithian@stanford.edu}\thanksref{tWill}}
\and
\author{\fnms{Trevor}~\snm{Hastie}\ead[label=e2]{hastie@stanford.edu}\thanksref{tTrev}}
\runauthor{W. Fithian and T. Hastie}
\affiliation{Stanford University}
\address{Department of Statistics\\
Stanford University\\
390 Serra Mall\\
Stanford, California 94305-4065\\
USA\\
\printead{e1}\\
\phantom{E-mail: }\printead*{e2}} 
\end{aug}
\thankstext{tWill}{Supported by NSF VIGRE Grant DMS-05-02385.}
\thankstext{tTrev}{Supported in part by NSF Grant DMS-10-07719 and NIH
Grant RO1-EB001988-15.}

\received{\smonth{6} \syear{2013}}
\revised{\smonth{3} \syear{2014}}

%
\begin{abstract}
For classification problems with significant class imbalance,
subsampling can reduce computational costs at the price of
inflated variance in estimating model parameters. We propose a
method for subsampling efficiently for logistic regression
by adjusting the class balance locally in feature space via
an accept--reject scheme. Our method generalizes standard
case-control sampling, using a pilot
estimate to preferentially select examples whose responses are
conditionally rare given their features. The biased subsampling is
corrected by a post-hoc analytic adjustment to the parameters. The
method is simple and requires one parallelizable scan over the
full data set.

Standard case-control sampling is inconsistent under
model misspecification for the population risk-minimizing
coefficients $\theta^*$. By contrast, our estimator
is consistent for $\theta^*$ provided that the pilot estimate is.
Moreover, under correct specification and with a consistent,
independent pilot estimate, our estimator has exactly twice
the asymptotic variance of the full-sample \mbox{MLE---}even if the
selected subsample comprises a miniscule fraction of the full data
set, as happens when the original data are severely imbalanced.
The factor of two
improves to $1+\frac{1}{c}$ if we multiply the baseline
acceptance probabilities by $c>1$ (and weight points with acceptance
probability greater than~1), taking roughly $\frac{1+c}{2}$
times\vspace*{1pt} as many data points into the subsample.
Experiments on simulated and real data show that our
method can substantially outperform standard case-control
subsampling.
\end{abstract}

%
\end{frontmatter}

\setcounter{footnote}{2}
\section{Introduction}\label{sec1}

In recent years, statisticians, scientists and engineers are
increasingly analyzing enormous data sets. When data
sets grow sufficiently large, computational costs may play a
major role in the analysis, potentially constraining our choice of
methodology or the number of data points we can afford to process.
Computational savings can
translate directly to statistical gains if they:
\begin{longlist}[(3)]
\item[(1)] enable us to experiment with and prototype
a variety of models, instead of trying only one or two,
\item[(2)] allow us to refit our models more often to adapt to changing
conditions,
\item[(3)] allow for cross-validation, bagging, boosting,
bootstrapping or other computationally intensive statistical
procedures or
\item[(4)] open the door to using more sophisticated statistical
techniques on a compressed data set.
\end{longlist}
\citet{bottou2008tradeoffs} discuss the tradeoffs arising when we
adopt this point of view. One simple manifestation of these
tradeoffs is that we may run out of computing resources before we run
out of data, in effect making the sample size $n$ a function of the
efficiency of our fitting method.

\subsection{Imbalanced data sets}

Class imbalance is pervasive in modern classification problems and
has received a great deal of attention in the machine learning
literature [\citet{chawla2004editorial}]. It can come in two forms:
\begin{description}
\item[] \textit{Marginal imbalance}. One of the classes is quite
rare; for instance, ${\mathbb{P}(Y\,{=}\,1)\approx0}$.
Such imbalance typically occurs in data sets
for predicting click-through rates in online advertising,
detecting fraud or diagnosing rare diseases.
\item[] \textit{Conditional imbalance}. For most values of the features
$X$, the response $Y$ is very easy to predict; for instance,
${\mathbb{P}(Y=1\gv X=0)\approx0}$ but $\mathbb{P}(Y=1\gv
X=1)\break \approx1$.
For example, such imbalance might arise in the
context of email spam filtering, where well-trained
classifiers typically make very few mistakes.
\end{description}
Both or neither of the above may occur in any given data set.
The machine learning literature on class imbalance usually
focuses on the first type, but the second type is also common.

If, for example, our data set contains one thousand or one million
negative examples for each positive example, then many of the negative data
points are in some sense redundant. Typically in such problems,
the statistical noise is primarily driven by the number of
representatives of the rare class, whereas the total size of the
sample determines the computational cost.
If so, we might hope to finesse our computational constraints
by subsampling the original data set in a way that
enriches for the rare class. Such a strategy must be implemented with
care if our ultimate inferences are to be valid for the full data set.

This article proposes one such data reduction scheme, local
case-control sampling, for use in fitting logistic regression models.
The method requires one parallelizable scan over the full data set and
yields a potentially much smaller subsample containing roughly half
of the information found in the original data set.

\subsection{Subsampling}

The simplest way to reduce the computational cost of a procedure is to
subsample the data before doing anything else.
However, uniform subsampling from an
imbalanced data set is inefficient since it fails to exploit the
unequal importance of the data points.

Case-control sampling---sampling uniformly
from each class but adjusting the mixture of the classes---is a more
promising approach. This procedure originated in epidemiology,
where the positive examples (cases) are typically diseased patients
and negative examples (controls) are disease-free
[\citet{mantel1959statistical}].
Often, an equal number of cases and controls are sampled, resulting in
a subsample with no marginal imbalance, and costly measurements of
predictor variables are only made for selected patients
[\citet{breslow1980statistical}].
This method is useful in our context as well, since
a logistic regression model fitted on the subsample can be converted
to a valid model for the original population via a simple adjustment
to the intercept [\citet{anderson1972separate,prentice1979logistic}].

However, standard case-control sampling still may not make most
efficient use of the data. For instance, it does nothing to exploit
conditional imbalance in a data set that is marginally balanced. Even
with some marginal imbalance, a control that looks similar to the cases
is often more useful for discrimination purposes
than one that is obviously not a case.

We propose a method, local case-control sampling, which attempts to
remedy imbalance \emph{locally} throughout the feature space. Given a
pilot estimate $(\tilde\alpha,\tilde\beta)$ of the logistic regression
parameters, local case-control sampling preferentially\vspace*{1pt} keeps
data points for which $Y$ is surprising
given $X$. Specifically, if $\tilde p(x) = \frac{e^{\tilde\alpha+
\tilde\beta'x}}{1+e^{\tilde\alpha+ \tilde\beta'x}}$, we accept
$(x_i,y_i)$ with probability $|y_i-\tilde p(x_i)|$, the $\ell_1$
residual of the pilot model.
In the presence of extreme marginal or conditional imbalance, these
errors will generally be quite small and the subsample can be many
orders of magnitude smaller than the full data set.

Just as with case-control sampling, we can fit our model to the
subsample and make an equally simple correction to obtain an estimate
for the original data set.
When the logistic regression model is correctly specified and the
pilot is consistent and independent of the data,
the asymptotic variance of the local case-control estimate is
exactly twice the variance of a logistic regression fit on the
(potentially much larger) full data set. This factor of two improves
to $1+\frac{1}{c}$ if we accept with probability $c|y_i-\tilde
p(x_i)|\wedge1$ and weight accepted points by a factor of
$c|y_i-\tilde p(x_i)|\vee1$. For example, if $c=5$ then the
variance of the subsampled estimate is only 20\% greater than the
variance of the full-sample MLE.
The subsample we take with $c>1$ is no more than $c$ times larger than
the subsample for $c=1$, and for data sets with large imbalance is
roughly $\frac{1+c}{2}$ times as large.

Local case-control sampling also improves on the bias of standard
case-control sampling. When the logistic regression model is
misspecified, case-control sampling is in general inconsistent for the
risk minimizer in the original population. By contrast, local
case-control sampling is always consistent given a consistent pilot,
and is also asymptotically unbiased when the pilot is.
Sections~\ref{secSimulations} and~\ref{secSpam} present empirical
results demonstrating the advantages of our approach in simulations and
on the Yahoo! webspam data set.

\subsection{Notation and problem setting}

Our setting is that of predictive classification: we
are given $n$ independent and identically distributed
observations, each consisting of predictors
$x_i\in\mathcal{X}$ and a binary response $y_i\in\{0,1\}$, with
joint probability measure $\mathbb{P}$.
For our purposes, we assume the predictors are mapped into some real
covariate vector space, so that $\mathcal{X}\subseteq\mathbb{R}^p$
Our aim
is to learn the function
%
\begin{equation}
p(x) = \mathbb{P}(Y=1|X=x)
\end{equation}
or equivalently to learn
%
\begin{equation}
f(x) = \logit\bigl(p(x)\bigr) = \log\frac{p(x)}{1-p(x)}
\end{equation}
which could be infinite for some $x$.

A linear logistic regression model assumes $f$ is linear in $x$; that is,
%
\begin{equation}
f_\theta(x) = f_{\alpha,\beta}(x) = \alpha+ \beta'x,
\end{equation}
where $\theta=(\alpha,\beta)\in\mathbb{R}^{p+1}$. This is less of
a restriction than it might
seem, since $x$ may represent a very large basis expansion of some
smaller set of ``raw'' features.

Nevertheless, in the real world, $f$ is unlikely to satisfy our
parametric model for any
given basis $x$. When the model is misspecified, we can still view
logistic regression as an M-estimator with convex loss equal to the
negative log-likelihood for a single data set:
%
\begin{equation}
\rho(\theta;x,y) = -y \bigl(\alpha+ \beta'x\bigr) + \log
\bigl(1+e^{\alpha
+\beta'x} \bigr).
\end{equation}

As an M-estimator, under general conditions logistic regression in
large samples will converge to the minimizer of the population risk
$R(\theta) = \mathbb{E}\rho(\theta;\break  X, Y)$ [\citet{huber2011robust}].
That is, $\theta$ converges to the population maximizer of the expected
log-likelihood
%
\begin{eqnarray}
\label{eqnPopRisk} \theta^* &=& \arg\min_\theta\mathbb{E}\rho(
\theta;X,Y)
\\
&=& \arg\min_\theta\mathbb{E} \bigl[-Y \bigl(\alpha+
\beta'X\bigr) + \log\bigl(1+e^{\alpha+\beta'X} \bigr) \bigr].
\end{eqnarray}

If $f=f_{\theta_0}$ for some $\theta_0$, then $\theta^*=\theta_0$;
otherwise $f_{\theta^*}$ is the best linear approximation to $f$ in
the sense of (\ref{eqnPopRisk}). For a misspecified model,
$f_{\hat\theta}$ cannot possibly converge to $f$ no matter
what sampling scheme or estimation procedure we use,
or how much data we obtain. Consistency, then, will mean that
$\hat\theta\stackrel{p}{\rightarrow}\theta^*$.

Model misspecification is ubiquitous in real-world
applications of regression methods. For reasons of exposition, the
misspecification always takes a simple form in our simulations, for
example, in Example~\ref{ex1} there are two binary predictors, and we would have
correct specification if only we added one interaction---but in the
real world it usually is neither possible nor even desirable to expand
the feature basis until the model is correctly specified. For
instance, if $p=1000$, then there are ${p+1\choose2}={}$500,500
quadratic terms. Even if we included all those terms as features, we
would still be missing cubic terms, quartic terms, and so on.

Some kinds of misspecification are milder than others, and some are
easier to find and fix than others. Seeking better-specified models
(without adding too much model complexity) is
a worthy goal, but realistically perfect specification is unattainable.

Our goal, then, is to speed up
computation while still obtaining a good estimate of $\theta^*$, the
population logistic regression parameters. As we will see, standard
case-control sampling achieves the first goal, but may fail
at the second.

\subsection{Related work}

Recent years have seen substantial work on classification in
imbalanced data sets.
See \citet{chawla2004editorial} and \citet{he2009learning}
for surveys of machine learning efforts on this
problem. Many of
the methods proposed involve some form of undersampling the
majority class, oversampling the minority class, or both.
\citet{owen2007infinitely} examined the limit of marginally imbalanced
logistic regression and proved it is equivalent to fitting an exponential
family model to the minority class.

One recurring theme is to preferentially sample
negative examples that lie near positive examples in feature space.
For example, \citet{mani2003knn} propose selecting majority-class
examples whose average distance to its three nearest minority examples
is smallest.
Our method has a similar flavor
since the probability of sampling a negative example $(x,0)$
is $\tilde p(x)$, which is large when the features $x$ are
similar to those characteristic of positive examples.

Our proposal lies more in the tradition of the epidemiological
case-control sampling literature. In particular, case-control
sampling within several categorical strata has been studied by
\citet{fears1986logistic,breslow1988logistic,weinberg1990design,scott1991fitting}.
Typically, the strata are based on easy-to-measure screening variables
available for a wide population,
with more laborious-to-collect variables being measured on the
sampled subjects. \citet{lumley2011connections} discuss survey
calibration methods for efficient regression in two-stage
sampling schemes, which are interesting but too computationally
intensive for our purposes here.

\section{Case-control subsampling}\label{secCC}

Case-control sampling is commonly carried out by taking all the cases
and exactly $c$ times as many controls for some fixed $c$ (e.g., $c =
1, 2, 5$). However, for our purposes it will be simpler
to consider a nearly equivalent procedure based on
accept--reject sampling.

Define some acceptance probability function $a(y)$ and let
$b=\log\frac{a(1)}{a(0)}$, the log-selection bias. Consider the
following algorithm:
\begin{longlist}[(3)]
\item[(1)] Generate independent $z_i\sim\operatorname{Bernoulli}(a(y_i))$.
\item[(2)] Fit a logistic regression to the subsample $S =
\{(x_i,y_i)\dvtx z_i=1\}$, obtaining unadjusted estimates
$\hat\theta_S=(\hat\alpha_S,\hat\beta_S)$.
\item[(3)] Assign
$\hat\alpha\gets\hat\alpha_S - b$ and $\hat\beta\gets\hat\beta_S$.
\end{longlist}
Specifically, we could generate the $z_i$ by first generating
$u_i\sim U(0,1)$ mutually independent of the pilot, the data, and each
other, then taking $z_i=\mathbf{1}_{u_i\leq a(y_i)}$. Note that steps
(2)--(3) are
equivalent to logistic regression with offset $b$ for each data point.

This variant is convenient to analyze because
the subsample thus obtained is an i.i.d. sample from a
new population:
%
\begin{equation}
\mathbb{P}_S(X,Y) = \mathbb{P}(X,Y\gv Z=1) = \frac{a(Y)\mathbb
{P}(X,Y)}{\bar a}
\end{equation}
with $\bar a = a(1)\mathbb{P}(Y=1) + a(0)\mathbb{P}(Y=0)$,
the marginal probability of $Z=1$.

The estimate $(\hat\alpha,\hat\beta)$ is motivated by a simple application
of Bayes' rule relating the odds of $Y=1$ in $\mathbb{P}$ and $\mathbb
{P}_S$. If
$g(x)$ is the true conditional log-odds function
for~$\mathbb{P}_S$, we have
%
\begin{eqnarray}
\label{eqnBayesCCBegin} g(x) &=& \log\frac{\mathbb{P}(Y=1\gv
X=x,Z=1)}{\mathbb{P}(Y=0\gv
X=x,Z=1)}
\\
&=& \log\frac{\mathbb{P}(Y=1\gv X=x)}{\mathbb{P}(Y=0\gv X=x)} + \log\frac
{\mathbb{P}(Z=1\gv Y=1,X=x)}{\mathbb{P}(Z=1\gv
Y=0,X=x)}
\\
\label{eqnBayesCCEnd} &=& f(x) + b.
\end{eqnarray}
That is, the log-odds $g(x)$ in our biased population is simply a
vertical shift by $b$ of the log-odds $f(x)$ in the original
population, so given an estimate of $g$ we can subtract $b$ to
estimate $f$.
If the model is correctly specified, logistic regression
on the subsample yields a consistent estimate for the function $g(x)$,
so the estimate for $f(x)$ is also consistent.

Note that the derivation (\ref{eqnBayesCCBegin})--(\ref{eqnBayesCCEnd})
is equally valid if the sampling bias $b$ depends on $x$, in which
case we have $g(x) = f(x) + b(x)$. Local case-control sampling
exploits this more general identity.

\subsection{Conditional probability and the logit loss}\label{subsecLogitLoss}

If $X$ is integrable, then upon differentiating the population risk
(\ref{eqnPopRisk}) with respect to $\theta$ we obtain the population
score criterion:
%
\begin{eqnarray}
\label{eqnPopScore} 0 &=& \mathbb{E} \biggl[ \biggl(Y-\frac{e^{f_\theta
(X)}}{1+e^{f_\theta
(X)}} \biggr)
\pmatrix{1
\cr
X} \biggr] = \int{ \bigl(p(x) - p_\theta(x) \bigr)
\pmatrix{1
\cr
x} \,d\mathbb{P}(x)}.
\end{eqnarray}
Informally, the best linear predictor is the one that gets
the conditional probabilities right on average. Note
this is not the same as a predictor that gets the conditional
log-odds right on average.

\begin{figure}

\includegraphics{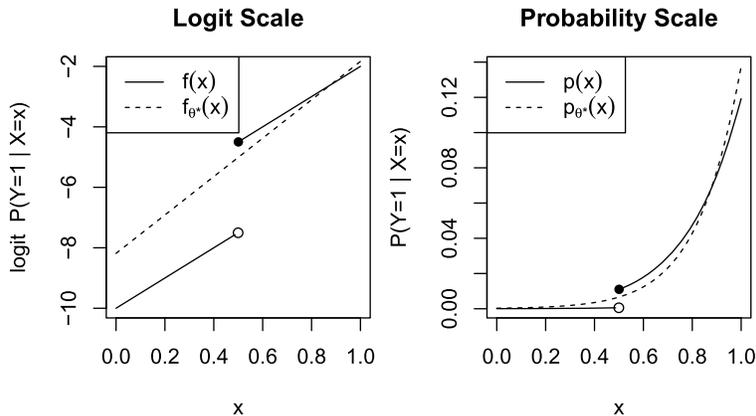}

\caption{The best linear fit $f_{\theta^*}(x)$
approximates the true log-odds $f(x)$ in the sense of matching
its implied conditional probabilities, not logits.}\label{figJumpPlots}
\end{figure}

To illustrate the difference between
approximating probabilities and approximating logits, suppose that
$X\sim
U(0,1)$ and $f(x) = -10 + 5x + 3\cdot\mathbf{1}_{x>0.5}$. The left
panel of
Figure~\ref{figJumpPlots} shows $f(x)$ as a solid line and its best linear
approximation as a dashed line. On the logit scale, the dashed line
appears to be
a very poor fit to the black curve. It fits reasonably well for
large $x$, but it appears more or less to ignore the smaller values of $x$.

The right panel of Figure~\ref{figJumpPlots} shows why. When we
transform both curves to the
probability scale, the fit looks much more reasonable.
$f_{\theta^*}(x)$ need not approximate $f(x)$ particularly well for
small $x$, because in that range even a large change in the log-odds
produces a negligible change in the conditional probability $p(x)$.
By contrast, $f_{\theta^*}(x)$ needs to approximate $f(x)$ well
for larger $x$, where $p(x)$ changes more
rapidly.
%

In general, logistic regression places highest priority on
fitting $f$ where $\frac{d p(x)}{d f(x)}$ is largest: where
$f(x)\approx0$ and $p(x)\approx0.5$.
In this example, with its strong marginal imbalance, the regions
that matter most are those where $p(x)$ is largest. This often makes
sense in applications such as medical screening or
advertising click-through rate prediction, where accuracy is most
important when the probability of
disease or click-through is nonnegligible.
In Section~\ref{secDiscussion}, we consider how to modify the
method to obtain classifiers that prioritize correctness near
some other, user-defined level curve of $p(x)$.

Finally, note that Figure~\ref{figJumpPlots} suggests the case-control
sampling estimate is unlikely to be
consistent for $\theta^*$ in general.
The nature of our linear
approximation in the left panel is intimately related to the fact that
$f(x) < 0$ everywhere in the sample space. If $f(x)$ were shifted
upward by some constant, the response of the dashed curve would be
more complicated than a simple constant shift by $b$, since the
relative importance of the two segments would change. Therefore,
estimating $f(x)+b$ and then subtracting $b$ may not be a successful
strategy.

\subsection{Inconsistency of case-control under misspecification}\label{subsecCCBias}

If the linear model is misspecified, the case-control estimate is
generically not consistent for the best linear predictor
$\theta^*$ as $n\rightarrow\infty$ [\citet
{xie1989logit,manski1989estimation}].
The unadjusted estimate will instead
converge to the best linear predictor of $g$ for the distribution
$\mathbb{P}_S$, which solves the score criterion
%
\begin{equation}
\label{eqnPopScoreCCUnadj} 0 = \int{ \biggl(\frac
{e^{f(x)+b}}{1+e^{f(x)+b}}-\frac{e^{f_\theta
(x)}}{1+e^{f_\theta(x)}} \biggr)
\pmatrix{1
\cr
x} \,d\mathbb{P}_S(x)}.
\end{equation}
Let $\theta_{\mathrm{CC}}^*(b)$ be the large-sample limit of the \emph{adjusted}
case-control sampling estimate with bias $b$. Then $\theta_{\mathrm{CC}}^*(b)$
solves the population score criterion
%
\begin{equation}
\label{eqnPopScoreCC} 0 = \int{ \biggl(\frac
{e^{f(x)+b}}{1+e^{f(x)+b}}-\frac{e^{f_\theta
(x)+b}}{1+e^{f_\theta(x)+b}} \biggr)
\pmatrix{1
\cr
x} \,d\mathbb{P}_S(x)}
\end{equation}
which differs from (\ref{eqnPopScore}) in two ways. First, the
integral is
taken over a different distribution for $X$. Second, and more
importantly, the integrand is different. We are now approximating
$f(x)$ in a different sense than we were.

In general under misspecification, $\theta_{\mathrm{CC}}^*(b)$ is different
for every $b$. If we sample cases and controls equally, in the
limit we will get a different answer than if we sample twice as many
controls; and in either case we will get a different answer than if we
use the entire data set or subsample uniformly.

These differences can be quite consequential for our
inferences about $\beta$ or the predictive performance of our model,
as we see next.

\begin{ex}[(Oatmeal and disease risk)]\label{ex1}
In this fictitious example, we consider estimating the effect of
exposure to
oatmeal on a person's risk of developing some rare disease. Suppose
that $10\%$ of the population has a family history of the disease,
half the population eats oatmeal (independently of family history), and
that both exposure and family history are binary predictors. Suppose
further that the true conditional log-odds function $f(x)$ is given by
the top-left panel of Table~\ref{tabOatmeal}.


\begin{table}
\tablewidth=313pt
\caption{Disease risk in the full population, and in the population
created by case-control sampling with equal numbers in each class}\vspace*{6pt}\label{tabOatmeal}
  \begin{minipage}[b]{.45\textwidth}
    \centering
    \begin{tabular}{@{}lcc@{}}
      \multicolumn{3}{c}{\textit{Original population} ($\mathbb{P}$)}\\[6pt]
      \multicolumn{3}{c}{\textit{Conditional log-odds} ($f$)}\\
      \hline
      & \textbf{History} $\bolds{-}$ & \textbf{History} $\bolds{+}$\\
      \hline
      Oatmeal $-$ & \phantom{0}$-$5 & $-$4\\
      Oatmeal $+$ & $-$10 & $-$1\\
      \hline
    \end{tabular}
    \vskip+6pt
    \begin{tabular}{@{}lcc@{}}
      \multicolumn{3}{c}{\textit{Conditional probabilities}}\\
      \hline
      & \textbf{History} $\bolds{-}$ & \textbf{History} $\bolds{+}$ \\
      \hline
      Oatmeal $-$ & 0.007 & 0.02\\
      Oatmeal $+$ & \phantom{.}5E$-$5 & 0.37\\
      \hline
    \end{tabular}
  \end{minipage}\quad
  \begin{minipage}[b]{.45\textwidth}
    \centering
    \begin{tabular}{@{}lcc@{}}
      \multicolumn{3}{c}{\textit{Case-control population} ($\mathbb{P}_S$)}\\[6pt]
      \multicolumn{3}{c}{\textit{Conditional log-odds} ($g$)}\\
      \hline
      & \textbf{History} $\bolds{-}$ & \textbf{History} $\bolds{+}$ \\
      \hline
      Oatmeal $-$ & $-$1.2 & $-$0.2 \\
      Oatmeal $+$ & $-$6.2 & \phantom{$-$}2.8 \\
      \hline
    \end{tabular}
    \vskip+6pt
    \begin{tabular}{@{}lcc@{}}
      \multicolumn{3}{c}{\textit{Conditional probabilities}}\\
      \hline
      & \textbf{History} $\bolds{-}$ & \textbf{History} $\bolds{+}$ \\
      \hline
      Oatmeal $-$ & 0.24\phantom{0} & 0.46\\
      Oatmeal $+$ & 0.002 & 0.94\\
      \hline
    \end{tabular}
  \end{minipage}\vspace*{-3pt}
\end{table}

The corresponding conditional probabilities $p(x)$ are given
in the lower-left panel of Table~\ref{tabOatmeal}.
Notice that oatmeal increases the risk for
people who are already at risk by virtue of their family history, but
has a protective effect for everyone else. This interaction means
that the additive logistic regression model is misspecified.

Because only the probabilities in the
``History $+$'' column are large enough to matter, the
fitted model for $f(x)$ pays more attention to the at-risk population,
for whom oatmeal elevates the risk of disease. A
logistic regression on a large sample from this population estimates
the coefficient for oatmeal as $\beta_{\mathrm{Oatmeal}}^* = 1.4$, implying
an odds ratio of about 4.0. This is
close to the marginal odds ratio of roughly 4.3 that we would obtain
if we did not control for family history.

Suppose, however, that we sampled an equal number of cases and
controls. Then the conditional log-odds of disease in our sample
would reflect the top-right panel of Table~\ref{tabOatmeal}, with all cells
increased by the same amount.

For large samples, the case-control estimate is $\beta_{\mathrm
{CC},\mathrm{Oatmeal}}^* =
-0.83$, implying an odds ratio of about 0.44. Using case-control
sampling has reversed our inference about the effect of oatmeal
exposure, because after
shifting the log-odds the left column becomes
much more important.
\end{ex}

\begin{figure}

\includegraphics{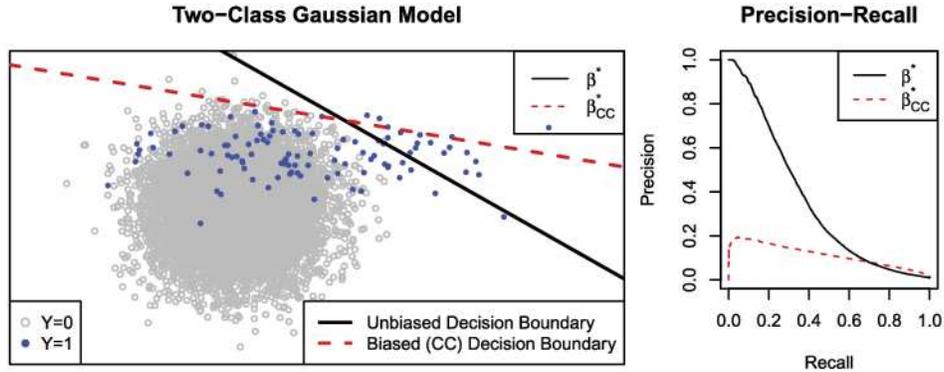}

\caption{At left, biased (case-control) and unbiased decision
boundaries for the bivariate Gaussian mixture model. At right, precision--recall
curves for $\beta^*$ and $\beta_{\mathrm{CC}}^*$.}\label{fig2DMix}
\end{figure}

\begin{ex}[(Two-class Gaussian model)]\label{ex2}
Suppose that $\mathbb{P}(Y=1)=1\%$, and that $X\gv Y \sim
\mathrm{N} (\mu_Y,\Sigma_Y )$. Let
%
\begin{eqnarray}
\mu_0&=&(0,0), \qquad \Sigma_0=\pmatrix{1&0\cr 0&1},
\\
\mu_1&=&(1.5,1.5),\qquad \Sigma_1=\pmatrix{0.3&0\cr 0&5}.
\end{eqnarray}
Data simulated from this model are shown in
the left panel of Figure~\ref{fig2DMix}. In this example, the
true log-odds $f(x)$ is an additive quadratic function of the two
coordinates $X_1$ and $X_2$.

In this example as in the previous one, the population-optimal
case-control parameters differ substantially from the optimal
parameters in the original population, with dramatic effects for the
predictive performance of the model. The \mbox{decision} boundaries for the two
estimates are overlayed on the left panel of Figure~\ref{fig2DMix}.
In the right panel, we plot the precision--recall
curves resulting from each set of parameters on a large test set.
\end{ex}

\subsection{Weighted case-control sampling}\label{subsecWCC}
A simple alternative to standard case-control sampling is to weight
the subsampled data points by the inverse of their probability of being
sampled.
We include weighted case-control sampling as a
competitor in our simulation studies in Section~\ref{secSimulations}.
Because it is a Horvitz--Thompson estimator with positive sampling
probabilities for any $(x,y)$ pair,
this method is $\sqrt{n}$-consistent, and asymptotically normal and
unbiased under general conditions [\citet{horvitz1952generalization}].

Although weighting succeeds in removing the bias induced by
the case-control sampling, this consistency comes at a cost of
increasing the variance, since the effective sample size is reduced
[\citeauthor{scott1986fitting} (\citeyear{scott1986fitting,scott2002robustness})].

%

Despite its inefficiency, the weighted case-control method can be an
attractive means of obtaining a consistent pilot if another good pilot
is not immediately available, and we later will use it to that end
in our experiments.

\section{Local case-control subsampling}\label{secLCC}

In this section, we describe local case-control subsampling, a
generalization of standard case-control sampling that both improves
on its efficiency and resolves its problem of inconsistency. To
achieve these benefits, we require a pilot estimate, that is, a good
guess $\tilde\theta= (\tilde\alpha,\tilde\beta)$ for the
population-optimal $\theta^*$.

\subsection{The local case-control sampling algorithm}

Local case-control sampling differs from case-control sampling only
in that the acceptance probability $a$ is allowed to depend on $x$ as
well as $y$. Our criterion for selection will be the degree of
``surprise'' we experience upon observing $y_i$ given $x_i$:
%
\begin{equation}
a(x,y) = \bigl|y-\tilde p(x)\bigr| = \cases{1-\tilde p(x), &\quad$y=1$,
\cr
\tilde p(x), &
\quad$y=0$,}
\end{equation}
where $\tilde p(x) = \frac{e^{\tilde\alpha+\tilde\beta
'x}}{1+e^{\tilde\alpha+\tilde\beta'x}}$ is the pilot
estimate of $\mathbb{P}(Y=1\gv X=x)$. The algorithm is:
\begin{longlist}[(3)]
\item[(1)] Generate independent $z_i\sim\operatorname{Bernoulli}(a(x_i,y_i))$.
\item[(2)] Fit a logistic regression to the sample $S=\{(x_i,y_i)\dvtx z_i=1\}$ to
obtain unadjusted estimates $\hat\theta_S = (\hat\alpha_S,\hat\beta_S)$.
\item[(3)] Assign $\hat\alpha\gets\hat\alpha_S + \tilde\alpha$ and
$\hat\beta\gets\hat\beta_S + \tilde\beta$.
\end{longlist}
As before,\vspace*{1pt} steps (2)--(3) are equivalent to fitting a logistic regression
in the subsample with offsets $-\tilde\alpha- \tilde\beta'x_i$.
The $z_i$ are generated as in Section~\ref{secCC}, and the adjustment
is again justified by
(\ref{eqnBayesCCBegin})--(\ref{eqnBayesCCEnd}), only now with
the constant $b$ replaced by
%
\begin{equation}
b(x) = \log\biggl(\frac{a(x,1)}{a(x,0)} \biggr) = -\tilde\alpha- \tilde
\beta'x.
\end{equation}
In other words, the subsample is drawn from a measure with
%
\begin{equation}
g(x) = f(x) - \tilde\alpha- \tilde\beta'x.
\end{equation}
If $f(x)$ is well approximated by the pilot estimate, then
$g(x) \approx0$ throughout feature space. That is,
conditional on selection into $S$, $y_i$ given $x_i$ is nearly a fair
coin toss.

To motivate this choice heuristically,
recall that the Fisher information for the log-odds of a
Bernoulli random variable is maximized when the probability is~$\frac{1}{2}$: fair coin tosses are more informative than
heavily biased ones.
In effect, local case-control sampling tilts the
conditional distribution of $Y$ given $X=x$ to make each $y_i$ in the
subsample more informative. We then fit a logistic regression in the
more favorable sampling measure, and ``tilt back''
to obtain a valid estimate for the original population.

In marginally imbalanced data sets where $\mathbb{P}(Y=1\gv X=x)$ is small
everywhere in the predictor space, a good pilot has
$\tilde p(x) \approx0$ for all $x$, and
the number of cases discarded by this algorithm will be quite small.
If we wish to avoid discarding any cases, we can always
modify the algorithm so that instead of keeping $(x,1)$ with
probability $a(x,1)$, we keep it with probability 1 and assign
weight $a(x,1)$.

\subsection{Choosing the pilot fit}

In many applications, there may be a natural choice of pilot fit
$\tilde\theta$; for instance, if we are refitting a classification
model every day to adapt to a changing world, then yesterday's fit is
a natural choice for today's pilot.

If no pilot fit is available from such a source, we recommend an
initial pass of weighted case-control sampling (described in
Section~\ref{subsecWCC}) to obtain the pilot. Weighted
case-control sampling using a fixed fraction of the original sample is
itself $\sqrt{n}$-consistent and asymptotically unbiased
for the true parameters. Consequently, if the pilot were fit using an
independent data set the second-stage estimate would
enjoy consistency and asymptotic unbiasedness
per the results in Section~\ref{secTheory}.

Our experiments suggest that mild inaccuracy in the pilot estimate,
and using a data-dependent pilot, do not unduly degrade the
performance of the local case-control algorithm. For example,
is Simulation~2 of Section~\ref{subsecSim2}, the pilot is fifty times
less efficient than the final local case-control estimate.
The main role of the pilot fit is to guide us in discarding most of
the data points for which $y_i$ is obvious given $x_i$ while keeping
those for which $y_i$ is conditionally surprising.

In our example and simulations, we use a pilot sample about the same
size as the local case-control subsample, on the principle that we
can afford to spend about as much time computing the pilot as computing the
second-stage estimate. When $\mathbb{P}(Y=1\gv X)$ is small throughout
$\mathcal{X}$,
this rule amounts roughly to weighted case-control sampling using all the
cases and one control per case. Although the above
rule has worked reasonably well
for us, at this time we can offer no finite-sample guarantees
that any given pilot sample size is large enough.

Because standard case-control sampling amounts to
local case-control sampling with a constant-only pilot fit, we might
expect that the pilot fit need not be perfect
to improve upon case-control sampling. Our experiments in
Sections~\ref{secSimulations} and~\ref{secSpam} support this
intuition.

\subsection{Taking a larger or smaller sample}\label{subsecLargerSample}

As we will see in Section~\ref{subsecAsymptotics}, under correct model
specification, and with an independent and consistent pilot,
the baseline procedure described above has exactly twice
the asymptotic variance as a logistic regression
estimated with the full sample, despite using a potentially very
small subset of the data. We can improve upon
this factor of two by increasing the size of the subsample.

One simple way to achieve this is to multiply all acceptance
probabilities by some constant $c$, for example, $c=5$. When deciding whether
to sample the point $(x_i,y_i)$, we would then generate $z_i\sim
\operatorname{Bernoulli}(ca(x_i,y_i)\wedge1)$ and assign weight
$w_i=ca(x_i,y_i)\vee1$ to each sampled point. This amounts to a
larger, weighted subsample from $\mathbb{P}_S$, and we can make the same
correction to the estimates from the subsample. We see in
Section~\ref{subsecAsyLargerSample} that for $c>1$ the factor of two
is replaced by a factor of $1+\frac{1}{c}$.\vspace*{1pt}

In the case of large imbalance, most of the $\tilde p(x_i)$ are near 0
or 1. For $c>1$, the marginal acceptance probability at $x_i$ becomes
%
\begin{eqnarray}
\mathbb{P}(z_i=1\gv x_i=x) &=& p(x) \bigl(c\bigl(1-
\tilde p(x)\bigr) \wedge1\bigr) + \bigl(1-p(x)\bigr) \bigl(c\tilde
p(x)\wedge1
\bigr)
\\
&\approx&(1+c)p(x) \bigl(1-p(x)\bigr),
\end{eqnarray}
where the approximation holds for $p(x)\approx\tilde p(x) \approx0$
or 1. For $c=1$, the marginal acceptance probability is
$p(x)(1-\tilde p(x)) + (1-p(x))\tilde p(x) \approx2p(x)(1-p(x))$, so
for $c>1$ we take roughly $\frac{1+c}{2}$ times as many data points as
for $c=1$.
For example, if $c=5$, the subsample accepted is roughly 3 times as
large, and the relative efficiency improves from $1/2$ to $5/6$.

Alternatively, if $n$ is extremely
large, even a small fraction of the
full data set may still be more than we want. In that case, we can
proceed as above with $c<1$, or simply sample any desired number $n_s$
of data points uniformly from the local case-control subsample.

\section{Asymptotics of the local case-control estimate}\label{secTheory}

We now turn to examining the asymptotic behavior of the local
case-control estimate. We first establish consistency, assuming a
consistent pilot estimate
$\tilde\theta$. We expressly do not assume that the pilot estimate is
independent of the data, since in some cases we may recycle
into the subsample some of the data we used to calculate the pilot.

By assuming independence of $\tilde\theta$ and the data,
we can obtain finer results about the
asymptotic distribution of $\hat\theta$. We show it is
asymptotically unbiased when $\tilde\theta$ is, and derive the
asymptotic variance of the estimate. When the logistic regression
model is correctly specified, the local case-control estimate
has exactly twice the asymptotic
variance of the MLE for the full data set.


\subsection{Preliminaries}

For better clarity of notation in this section, we will use the
letter $\lambda$ in place of $\tilde\theta$ to denote
pilot estimates. Additionally, we drop the notation ${1\choose x}$ and
absorb the constant term into $x$, so that
$f_\theta(x)=\theta'x$.
To avoid trivialities, assume without loss of generality that there
is no $v\in\mathbb{R}^p$ for which $\mathbb{E}|v'X|=0$ (if not, we can
discard redundant features).

For $\pi\in[0,1]$ define the ``soft hinge'' function
%
\begin{equation}
h(\eta; \pi) = - \pi\eta+ \log\bigl(1+e^\eta\bigr),
\end{equation}
and note that
%
\begin{equation}
\mathbb{E} \bigl[\rho(\theta; X,Y) \gv X=x \bigr] = h\bigl(\theta'x;
p(x)\bigr).
\end{equation}
As a function of $\eta$, $h$ is positive and strictly convex, its
magnitude is
bounded by ${1+|\eta|}$, and it has Lipschitz constant ${\max(\pi,1-\pi)
\leq1}$.
If $\pi<1$, $h$ diverges as $\eta\rightarrow\infty$, and if $\pi>0$
$h$ diverges as $\eta\rightarrow-\infty$.\vadjust{\goodbreak}

As a function of $\lambda$,
$a_\lambda(x,y)=\llvert y-\frac{e^{\lambda'x}}{1+e^{\lambda
'x}}\rrvert\in
(0,1)$ has Lipschitz constant $\leq\|x\|$.
Hence, $\bar a(\lambda)=\mathbb{E}a_\lambda(X,Y)\in(0,1)$ with
Lipschitz constant $\leq\mathbb{E}\|X\|$.
The marginal acceptance probability given $x$ is
%
\begin{equation}
\hat a_\lambda(x) = \tilde p(x) \bigl(1-p(x)\bigr) + \bigl(1-\tilde p(x)
\bigr)p(x) 
\in(0,1).
\end{equation}

Given pilot $\lambda$, the local case-control subsampling scheme
effectively samples from the probability measure $\mathbb{P}_\lambda
$, where
%
\begin{equation}
d\mathbb{P}_\lambda(x,y)=\frac{a_\lambda(x,y)\,d\mathbb
{P}(x,y)}{\bar a(\lambda)},
\end{equation}
and $\bar a(\lambda) = \int a_\lambda(x,y) \,d\mathbb{P}(x,y)$
is the marginal
probability of acceptance. Under this measure,
%
\begin{equation}
\logit\mathbb{P}_\lambda(Y=1\gv X=x) = f(x)-\lambda'x.
\end{equation}
Because $a_\lambda(x,y)\leq1$, if $g(X,Y)$ is integrable under
$\mathbb{P}$ it
is also integrable under any $\mathbb{P}_\lambda$.

Conditioning on $X$, we can write the population risk of the logistic
regression parameters $\theta$ with respect to sampling measure
$\mathbb{P}_\lambda$ as
%
\begin{equation}
\label{eqnPopRiskLam2} R_\lambda(\theta) = \frac{-1}{\bar a(\lambda)}
\int{ h \biggl(
\theta'x; \frac{e^{f(x)-\lambda'x}}{1+e^{f(x)-\lambda
'x}} \biggr) \hat a_\lambda(x) \,d
\mathbb{P}(x)}.
\end{equation}

By Cauchy--Schwarz, the integrand in (\ref{eqnPopRiskLam2}) is bounded by
$2(1+\|\theta\| \|x\|)$. If $\mathbb{E}\|X\|<\infty$, then, we may
appeal to dominated convergence and
take limits with respect to $\theta$ and $\lambda$ inside the integral.

$R_\lambda(\theta)$ is strictly convex because the integrand is, and
always has a unique population minimizer if there is no separating hyperplane
in the population.

\begin{lemma}\label{lemma1}
Assume there is no $v$ for which
%
\begin{equation}
\mathbb{P}\bigl(Y=0, v'X>0\bigr) = \mathbb{P}\bigl(Y=1,v'X<0
\bigr) = 0.
\end{equation}
Henceforth, we refer to this assumption as \emph{nonseparability}.
Then $R_\lambda(\theta)$ attains a unique minimum for
every $\lambda\in\mathbb{R}^p$.
\end{lemma}

Denote by $\Rhat_{\lambda}^{(0)}(\theta)$ the empirical risk on a
local case-control subsample taken using the pilot estimate $\lambda$.
Then
%
\begin{equation}
\label{eqnEmpRisk0} \Rhat_{\lambda}^{(0)}(\theta) = - \Biggl(\sum
_{i=1}^n{z_i}
\Biggr)^{-1}\sum_{i=1}^nz_i
\bigl[y_i\theta'x_i - \log
\bigl(1+e^{\theta'x_i} \bigr) \bigr].
\end{equation}
It will be somewhat simpler to replace the
random subsample size $\sum_{i=1}^n{z_i}$
with its expectation $n\bar a(\lambda)$. Define
%
\begin{equation}
\label{eqnEmpRisk1} \Rhat_{\lambda}(\theta) = -\frac{1}{n\bar a(\lambda
)}\sum
_{i=1}^nz_i \bigl[y_i
\theta'x_i - \log\bigl(1+e^{\theta'x_i} \bigr)
\bigr].
\end{equation}
Since minimizing (\ref{eqnEmpRisk0}) with respect to $\theta$ is
equivalent to
minimizing (\ref{eqnEmpRisk1}), the two are equivalent for our purposes.

If the unadjusted parameters $\hat\theta_S$ minimize
$\Rhat_\lambda$, the local case-control
estimate $\hat\theta=\hat\theta_S+\lambda$ is an $M$-estimator
minimizing $\Qhat_\lambda(\theta)=\Rhat_\lambda(\theta-\lambda)$.
We use analogous notation for the population version:
%
\begin{equation}
Q_\lambda(\theta) = R_\lambda(\theta-\lambda).
\end{equation}
For any given pilot estimate $\lambda$ and large $n$, we expect
%
\begin{equation}
\label{eqnArgMinQLam} \hat\theta\approx\arg\min_{\theta}
Q_{\lambda}(\theta).
\end{equation}
Define the right-hand side of (\ref{eqnArgMinQLam}) to be
$\bar\theta(\lambda)$, the large-sample limit of local case-control
sampling with pilot estimate fixed at $\lambda$. The best linear
predictor for the original population corresponds to the case
$\lambda=0$ (uniform subsampling), that is, $\theta^*=\bar\theta(0)$.
Consistency means that for large $n$, $\hat\theta\stackrel
{p}{\rightarrow}\theta^*$.

Recall that if the model is correctly specified with true
parameters $\theta_0$, then $\bar\theta(\lambda) = \theta_0$
for \emph{any} fixed pilot estimate $\lambda$.
Minimizing $\Qhat_\lambda$ therefore yields a \mbox{consistent} estimate.
Unfortunately, in the misspecified case $\bar\theta(\lambda)
\neq
\bar\theta(0)=\theta^*$. In this
sense, local case-control sampling with the pilot $\lambda$ held fixed
is in
general \emph{not}~consistent for $\theta^*$.
However, we see below that it is consistent if $\lambda=\theta^*$.
%
\begin{proposition}\label{propThetaStarThetaStar}
Assume $\mathbb{E}\|X\|<\infty$, that the classes are nonseparable, and
that $\theta^*=\bar\theta(0)$ is the best linear predictor for the
original measure $\mathbb{P}$. Then
%
\begin{equation}
\theta^* = \arg\min_\theta Q_{\theta^*}(\theta) = \bar
\theta\bigl(\theta^*\bigr).
\end{equation}
\end{proposition}
In other words, if we could only choose our pilot perfectly, then the
local case-control estimate would converge to $\theta^*$ as
$n\rightarrow\infty$.

\begin{pf*}{Proof of Proposition \ref{propThetaStarThetaStar}}
Write $p^*(x) = \frac{e^{{\theta^*}'x}}{1+e^{{\theta^*}'x}}$.
The population optimality criterion for LCC with pilot $\lambda$ is
%
\begin{eqnarray}
0 
&=& -\bar a(\lambda) \nabla_\theta R_\lambda(\theta-
\lambda)
\\
&=& -\mathbb{E} \bigl[X\rho'\bigl((\theta-\lambda)'X;
X, Y\bigr) a_\lambda(X,Y) \bigr].
\end{eqnarray}
Noting that $-\rho'(0; x, y) = y-\frac{1}{2}$, if we evaluate the above
at $\lambda=\theta=\theta^*$, we obtain
%
\begin{eqnarray}
\label{eqnPopOptQlam} \mathbb{E} \bigl[X \bigl(Y-\tfrac{1}{2} \bigr)
a_\lambda(X,Y) \bigr] &=& \tfrac{1}{2}\mathbb{E} \bigl[ X \bigl(p(X)
\bigl(1-p^*(X)\bigr) - \bigl(1-p(X)\bigr)p^*(X) \bigr) \bigr]\hspace*{-25pt}
\\
&=& \tfrac{1}{2}\mathbb{E} \bigl[ X \bigl(Y-p^*(X) \bigr) \bigr]
\end{eqnarray}
which is exactly half the population score (\ref{eqnPopScore}) for the
original population. Since $\theta^*$ optimizes the risk for the
original population, this value is 0.\vadjust{\goodbreak}
\end{pf*}


There is an intuitive explanation of this result: in $\mathbb
{P}_{\theta^*}$,
the \emph{acceptance probabilities} are $p^*(X)$ if $Y=0$ and $1-p^*(X)$ if
$Y=1$; hence they play the same role as the \emph{pseudoresiduals}
$Y-p^*(X)$ did in the original measure $\mathbb{P}$. For example, the point
$(x,0)$ would contribute $p^*(x) x$ to the gradient if we evaluated
the full-sample score at $\theta^*$. Evaluating the subsample score
at 0, the same point now contributes $\frac{1}{2}x$ to the
score---but only if it is accepted, which occurs with probability
exactly $p^*(x)$. So, in essence, the subsampling stands in for the
reweighting that we otherwise would have done when fitting our
logistic regression to the full sample.

Of course, in practice we never have a perfect pilot---if we did we would not need to estimate $\theta^*$---but
Proposition~\ref{propThetaStarThetaStar}
suggests that if $\lambda$ is near $\theta^*$,
minimizing $\Qhat_\lambda$ yields a good estimate. In fact, we will
see that if
$\lambda\stackrel{p}{\rightarrow}\theta^*$ then
$\hat\theta\stackrel{p}{\rightarrow}\theta^*$ as well.

\subsection{Consistency}

For our asymptotic results, assume we have an infinite reservoir
$(x_1,y_1), (x_2,y_2), \ldots$ of i.i.d. pairs, a sequence of
i.i.d. $U(0,1)$ variables $u_1, u_2,\ldots$ for making accept--reject
decisions, and a sequence of pilot estimates
$\lambda_1,\lambda_2,\ldots.$ The $\lambda_n$ are possibly dependent
upon the data, but the $u_i$ are assumed to be independent of
everything else.

$\hat\theta_n$ is the local case-control estimate, computed using
pilot $\lambda_n$, data\break  $\{(x_i,y_i)\}_{i=1}^n$, and accept--reject
decisions $z_i = \mathbf{1}_{u_i\leq a_{\lambda_n}(x_i,y_i)}$.

The main result of this section is that if the pilot estimate
$\lambda_n$ is consistent for $\theta^*$, then so is $\hat\theta_n$.
The details are somewhat technical, especially the proof of
Proposition~\ref{prop_pointwise}, but the main idea is that if
$\lambda_n\stackrel{p}{\rightarrow}\theta^*$, then for large $n$
%
\begin{equation}
\Qhat_{\lambda_n} \approx Q_{\theta^*}
\end{equation}
in the appropriate sense. $\Qhat_{\lambda_n}$ is what the local
case-control estimate actually minimizes, whereas the last function is
minimized by $\theta^*$, our ultimate target.

First, we establish pointwise convergence.
%
\begin{proposition}\label{prop_pointwise}
If $\mathbb{E}\|X\| <\infty$ and
$\lambda_n\stackrel{p}{\rightarrow}\lambda_\infty$, then for each
$\theta\in\mathbb{R}^p$,
%
\begin{equation}
\label{eqnPointwiseConv} \Qhat_{\lambda_n}(\theta)\stackrel{p}
{\rightarrow}Q_{\lambda
_\infty}(
\theta).
\end{equation}
\end{proposition}
Because we avoid assuming independence between the
pilot $\lambda_n$ and the data $(x_i,y_i)$, the proof is
technical and is deferred to the \hyperref[appe]{Appendix}. The proof relies on
the coupling of the acceptance decisions $z_i$ for different pilot estimates
through $u_i$. With this coupling, two
nearby pilot estimates will differ on very few accept--reject
decisions.

Because neither $\Qhat_{\lambda_n}(\theta)$ nor
$Q_{\lambda_\infty}(\theta)$ changes very fast, pointwise convergence
also implies uniform convergence on compacts.
%
\begin{proposition}\label{propUniform}
If $\mathbb{E}\|X\| <\infty$ and $\lambda_n\stackrel{p}{\rightarrow
}\lambda_\infty$, then for compact $\Theta\subseteq\mathbb{R}^p$,
%
\begin{equation}
\sup_{\theta\in\Theta} \bigl|\Qhat_{\lambda_n}(\theta)-Q_{\lambda_\infty}(
\theta)\bigr|\stackrel{p} {\rightarrow}0.\vadjust{\goodbreak}
\end{equation}
\end{proposition}

\begin{pf}
Define
%
\begin{equation}
F_n(\theta) = \Qhat_{\lambda_n}(\theta)-Q_{\lambda_\infty}(\theta).
\end{equation}
By Proposition~\ref{prop_pointwise}, $F_n(\theta)\stackrel
{p}{\rightarrow}0$ pointwise. Next, we
show it is Lipschitz.
The integrand in (\ref{eqnPopOptQlam}) is $x$ times two factors
each bounded by $\pm1$, hence
%
\begin{equation}
\bigl\|\bar a(\lambda_\infty)\nabla_\theta Q_{\lambda_\infty}\bigr\| \leq
\int{\|x\| \,d\mathbb{P}(x)} = \mathbb{E}\|X\|.
\end{equation}
Similarly for $\Qhat_{\lambda_n}$, we have
%
\begin{equation}
\nabla_\theta\Qhat_{\lambda_n} = -\frac{1}{n\bar a(\lambda_n)} \sum
_{i=1}^n z_i \biggl(y_i-
\frac{e^{(\theta-\lambda
_n)'x_i}}{1+e^{(\theta-\lambda_n)'x_i}} \biggr)x_i
\end{equation}
so that
%
\begin{equation}
\sup_\theta\|\nabla_\theta\Qhat_{\lambda_n}\| \leq
\bar a(\lambda_n)^{-1} \frac{1}{n}\sum
_{i=1}^n \|x_i\| \stackrel{p} {
\rightarrow}\bar a(\lambda_\infty)^{-1}\mathbb{E}\|X\|.
\end{equation}
It follows that, with probability tending to 1, $F_n(\theta)$ has
Lipschitz constant less than $c = 3\bar a(\lambda_\infty
)^{-1}\mathbb{E}\|X\|$.

Now, for any $\varepsilon>0$, we can cover $\Theta$ with finitely many
Euclidean balls of radius $\delta=\varepsilon/c$, centered at
$\theta_1,\ldots,\theta_{M(\varepsilon)}$. Let $A_n(\varepsilon)$
be the event that
$F_n$ has Lipschitz constant less than $c$ and
%
\begin{equation}
\sup_{1\leq j\leq M(\varepsilon)} \bigl|F_n(\theta_j)\bigr| <
\varepsilon.
\end{equation}
On $A_n(\varepsilon)$, we have $\sup_{\theta\in\Theta} |F_n(\theta
)| < 2\varepsilon$,
and $\mathbb{P}(A_n(\varepsilon))\rightarrow1$ as $n\rightarrow
\infty$.
\end{pf}

Finally, we come to the main result of the section, in which we prove
that the local case-control estimate is consistent when the pilot is.
Because the functions are strictly convex, we can ignore everything but
a neighborhood of $\theta^*$.

%
\begin{theorem}\label{thmConsistency}
Assume $\mathbb{E}\|X\| < 0$ and the classes are nonseparable.

If $\lambda_n\stackrel{p}{\rightarrow}\theta^*$ then the local
case-control estimate
$\hat\theta_n\stackrel{p}{\rightarrow}\theta^*$ as well.
\end{theorem}

\begin{pf}
Let $\Theta\subseteq\mathbb{R}^p$ be any compact set with $\theta
^*$ in its
interior, and let
%
\begin{equation}
\varepsilon= \inf_{\theta\in\partial\Theta}Q_{\theta^*}(\theta
)-Q_{\theta
^*}
\bigl(\theta^*\bigr) > 0,
\end{equation}
where the strict inequality follows from strict convexity. Uniform
convergence implies that with probability tending to 1,
%
\begin{equation}
\sup_{\theta\in\Theta}\bigl|\Qhat_{\lambda_n}(\theta)- Q_{\theta^*}(
\theta)\bigr| < \varepsilon/2
\end{equation}
which implies in turn that
%
\begin{equation}
\inf_{\theta\in\partial\Theta}\Qhat_{\lambda_n}(\theta) > \Qhat
_{\lambda_n}
\bigl(\theta^*\bigr).
\end{equation}
Whenever this is the case, the strictly convex function
$\Qhat_{\lambda_n}$ has a unique minimizer in the interior of
$\Theta$. Since $\Theta$ was arbitrary, we can take its diameter to
be less
than any $\delta>0$. Hence, $\hat\theta_n\stackrel{p}{\rightarrow
}\theta^*$.
\end{pf}

\subsection{Asymptotic distribution}\label{subsecAsymptotics}

In this section, we derive the asymptotic distribution of the
local case-control logistic regression estimate, in the same
asymptotic regime as the previous section. To prove our results
here, we assume the pilot estimate $\lambda_n$ is independent of our data
set. This would not be the case if our pilot were based on a
subsample of the data (the procedure we use for all our simulations),
but it could hold if the pilot came from a model fitted to data from
an earlier time period.

The main result of this section is that if the logistic regression
model is correctly specified and the pilot is consistent,
the asymptotic covariance matrix of the local case-control estimate for
$\theta$ is exactly twice the asymptotic covariance matrix of a logistic
regression performed on the entire data set.
For the results in this section, we will need $\mathbb{E}\|X\|
^2<\infty$.

It will be
convenient to give names to some recurring quantities. First, we have
seen that if $\mathbb{E}\|X\|<\infty$ we can differentiate $Q_\lambda
(\theta)$
inside the integral to obtain the gradient of the population risk:
%
\begin{eqnarray}\label{eqnIntegralG}
G(\theta,\lambda) &\triangleq& -\bar a(\lambda)\nabla_\theta
Q_\lambda(\theta)
\\
&=& \int{ \biggl(\frac{e^{f(x)-\lambda'x}}{1+e^{f(x)-\lambda'x}} - \frac
{e^{(\theta-\lambda)'x}}{1+e^{(\theta-\lambda)'x}} \biggr)
\hat a_\lambda(x) x \,d\mathbb{P}(x)}.
\end{eqnarray}
Whereas $G$ is the expectation of the logistic
regression score with respect to $\mathbb{P}_\lambda$, we can also
define its
covariance matrix:
%
\begin{equation}
J(\theta,\lambda) \triangleq\Var_\lambda\biggl[ \biggl(Y -
\frac{e^{(\theta-\lambda)'X}}{1+e^{(\theta-\lambda)'X}} \biggr) X \biggr].
\end{equation}
When $\mathbb{E}\|X\|^2<\infty$, $J(\theta,\lambda)<\infty$, and
is continuous
in $\theta$ and $\lambda$ by dominated convergence.\vadjust{\goodbreak}

Since the derivatives of the integrand in (\ref{eqnIntegralG})
are uniformly bounded by $2\|x\|^2$, dominated
convergence implies we can again differentiate inside the integral.
Differentiating with respect to $\theta$ we obtain
%
\begin{eqnarray}
H(\theta,\lambda) &\triangleq& -\bar a(\lambda)\nabla_\theta^2
Q_\lambda(\theta)
\\
&=& \int{ \frac{e^{(\theta-\lambda)'x}}{(1+e^{(\theta-\lambda)'x})^2}
\biggl(\frac{e^{\lambda'x}+e^{f(x)}}{(1+e^{\lambda
'x})(1+e^{f(x)})} \biggr) xx'
\,d\mathbb{P}(x)}.
\end{eqnarray}
Here, the integrand is dominated by $xx'$, so dominated convergence
again applies and thus we see that $H$ is continuous in $\theta$ and
$\lambda$. $H(\theta,\lambda)\succ0$ for
any $\theta, \lambda$ since we have assumed
there is no nonzero $v$ for which
$\mathbb{E}|v'X|=0$. Finally, define the matrix of crossed partials:
%
\begin{eqnarray}
C(\theta,\lambda) &\triangleq&\nabla_\lambda G(\theta,\lambda).
\end{eqnarray}
To be concrete, $C_{i,j} = \frac{\partial^2}{\partial\theta_i
\partial\lambda_j}
Q_\lambda(\theta)$. Continuity of $C$ again follows from noting the
derivative of the integrand in (\ref{eqnIntegralG}) with respect to
$\lambda$ is dominated by $8\|x\|^2$.

To begin, we consider the behavior of $\bar\theta(\lambda)$ for
$\lambda$ near $\theta^*$. By Proposition~\ref{propThetaStarThetaStar}, we have $G(\theta^*,\theta^*)=0$.
Since $H(\theta,\lambda)\succ0$,
we can apply the implicit function theorem to the relation
$G(\bar\theta(\lambda),\lambda)=0$ to obtain
%
\begin{equation}
\label{eqnCondlBias} \bar\theta(\lambda) = \theta^* + H\bigl(\theta
^*,\theta^*
\bigr)^{-1}C\bigl(\theta^*,\theta^*\bigr) \bigl(\lambda-\theta^*\bigr) +
o\bigl(\bigl\|\lambda-\theta^*\bigr\|\bigr).
\end{equation}

By standard M-estimator theory, if we fix $\lambda$ and send
$n\rightarrow\infty$ the coefficients of a logistic regression
performed on a sample of size $|S|$ from $\mathbb{P}_\lambda$ would be
asymptotically normal with covariance matrix
%
\begin{equation}
{\frac{1}{|S|}H\bigl(\bar\theta(\lambda),\lambda\bigr)^{-1} J
\bigl(\bar\theta(\lambda),\lambda\bigr)H\bigl(\bar\theta(\lambda
),\lambda
\bigr)^{-1}}.
\end{equation}

In light of this and the fact that $|S|\approx\bar a(\lambda)n$,
we might predict the following.

%
\begin{theorem}\label{thm_condl_var}
Assume $\mathbb{E}\|X\|^2<\infty$. If $\lambda_n\stackrel
{p}{\rightarrow}\theta^*$
independently of the data, then
%
\begin{equation}
\label{eqnCondlVar} \sqrt{n} \bigl(\hat\theta_n - \bar\theta(
\lambda_n) \bigr) \stackrel{\mathcal{D}} {\rightarrow} N\bigl(0,\bar a
\bigl(\theta^*\bigr)^{-1}\Sigma\bigr)
\end{equation}
with $\Sigma=
H(\theta^*,\theta^*)^{-1}J(\theta^*,\theta^*)H(\theta^*,\theta^*)^{-1}$.
\end{theorem}
Again, we defer the proof to the \hyperref[appe]{Appendix}. We can combine
(\ref{eqnCondlVar}) with (\ref{eqnCondlBias}) to immediately obtain the
following reassuring facts.
%
\begin{corollary}\label{corBiasVar}
Assume $\mathbb{E}\|X\|^2<\infty$ and $\lambda_n$ is a sequence of pilot
estimators given independently of the data. Then:
\begin{longlist}[(a)]
\item[(a)] If $\lambda_n$ is $\sqrt{n}$-consistent, so is $\hat\theta_n$.
\item[(b)] If $\lambda_n$ is asymptotically unbiased, so is
$\hat\theta_n$.\vadjust{\goodbreak}
\item[(c)] If $\sqrt{n} (\lambda_n-\theta^* ) \stackrel
{\mathcal{D}}{\rightarrow}N(0,V)$
then $\sqrt{n} (\hat\theta_n-\theta^* ) \stackrel
{\mathcal{D}}{\rightarrow}N (0,\Sigma)$ with
%
\begin{equation}
\label{eqnFullVar} \Sigma= H^{-1} \bigl(CVC' + \bar
a^{-1}J \bigr)H^{-1}.
\end{equation}
In (\ref{eqnFullVar}), we have suppressed the arguments of $\theta^*$
in $H,C,\bar a$ and $J$.
\end{longlist}
\end{corollary}
The first term in (\ref{eqnFullVar}) characterizes the contribution of
conditional bias (given~$\lambda_n$) to the overall variance, and
the second is the contribution of conditional variance.

In the special case where logistic regression model is correctly
specified, we have the following.
%
\begin{theorem}\label{thmAsyVar}
Assume the logistic regression model is correct and let
$\frac{1}{n}\Sigma_{\mathrm{full}}$ be the asymptotic variance of the
MLE for the full sample. Then if $\mathbb{E}\|X\|^2<\infty$ and
$\lambda_n\stackrel{p}{\rightarrow}\theta_0$ independently of the
data, we have
%
\begin{equation}
\label{eqnAsyVar2} \sqrt{n} (\hat\theta_n-\theta_0 )
\stackrel{\mathcal{D}} {\rightarrow}N \bigl(0,a(\theta_0)^{-1}
\Sigma\bigr) = N (0,2\Sigma_{\mathrm{full}} ).
\end{equation}
\end{theorem}

Hence, although
the size of a local case-control subsample is roughly $n\bar a(\lambda
)$, the variance of $\hat\theta$
is the same
as if we took a simple random sample of size $n/2$ from the full data set.
In other words, each point sampled is worth about
$\frac{1}{2\bar a(\lambda_n)}$ points sampled uniformly.

\begin{pf*}{Proof of Theorem \ref{thmAsyVar}}
If logistic regression is correctly specified for $\mathbb{P}$, it is also
for $\mathbb{P}_\lambda$, regardless of $\lambda$, so
$\bar\theta(\lambda)\equiv\theta_0$. Furthermore, by standard
maximum likelihood theory
$J(\theta_0,\lambda)=H(\theta_0,\lambda)^{-1}$ for each $\lambda$.
Therefore, (\ref{eqnCondlVar}) specializes to
%
\begin{equation}
\sqrt{n} (\hat\theta_n - \theta_0 ) \stackrel{\mathcal
{D}} {\rightarrow} N\bigl(0,\bar a(\theta_0)^{-1}H(
\theta_0,\theta_0)^{-1}\bigr).
\end{equation}
But
%
\begin{equation}
\label{eqnGeneralHessian} \qquad H(\theta,\lambda) = \bar a(\lambda)^{-1}\int{
\biggl[
\frac{e^{(\theta-\lambda)'x}}{(1+e^{(\theta-\lambda
)'x})^2} \biggr] \biggl[\frac{e^{\lambda'x}+e^{f(x)}}{(1+e^{\lambda
'x})(1+e^{f(x)})} \biggr] xx' \,d\mathbb{P}(x)}.
\end{equation}
If $f(x)=\theta_0'x$ and $\lambda=\theta_0$, then
(\ref{eqnGeneralHessian}) simplifies to
%
\begin{eqnarray}
H(\theta_0,\theta_0) &=& \bar a(\theta_0)^{-1}
\frac{1}{2} \int{\frac{e^{\theta_0'x}}{(1+e^{\theta_0'x})^2}xx'
\,d\mathbb{P}(x)}
\\
&=& \bar a(\theta_0)^{-1}\frac{1}{2} H(
\theta_0,0)
\\
&=& \bigl(2\bar a(\theta_0) \Sigma_{\mathrm{full}}
\bigr)^{-1}.
\end{eqnarray}\upqed
\end{pf*}

This result is surprisingly simple. No characterization like
Theorem~\ref{thmAsyVar} is available for the case-control and weighted
case-control estimates, whose variances are
not simple scalar multiples of $\Sigma_{\mathrm{full}}$.

We can offer a simple heuristic
argument for Theorem~\ref{thmAsyVar}, similar to that of
Proposition~\ref{propThetaStarThetaStar}. In $\mathbb{P}_{\theta
_0}$, the
acceptance probability $\hat a_\lambda(x)$ for an observation at $x$
is $2p(x)(1-p(x))$, and given that it is accepted it contributes
$\frac{1}{4}xx'$ to the observed information. In the full sample, a
point at $x$ is always accepted but contributes less,
$p(x)(1-p(x))xx'$, to the observed information. Again, the sampling
probability stands in for the reweighting we would have done in the
full sample.
If $p(x)(1-p(x))$ is very small,
we are discarding most of the data instead of keeping all of it and
assigning it a tiny weight in the fit.

The practical meaning of Theorem~\ref{thmAsyVar}
is that local case-control sampling is most advantageous when
$\bar a(\theta_0) = \mathbb{E}(|Y-\tilde p(X)|)$ is small, that
is, when $Y$
is easy to predict throughout much of the covariate space. This can
happen as a result of marginal or conditional imbalance, or both.
Standard case-control sampling can also improve our efficiency in the
presence of marginal imbalance, but unlike local case-control
sampling, it does not exploit conditional
imbalance. Hence, we would expect local case-control to
outperform standard case-control most dramatically
when the marginal imbalance is
very high, as in the simulation of Section~\ref{subsecSim2}.

For data-dependent pilots, the efficiency picture is somewhat more
complicated.~For example, $\bar\theta(\lambda)$ is
approximately a
linear function of $\lambda-\theta^*$.
Thus, if $\lambda$ is unbiased but correlated with the
noise in the data, we might get more or less variance relative
to~(\ref{eqnAsyVar2}), depending on how this correlation interacts
with $C$.
If the model is correctly specified, it less clear whether an
adversarially chosen pilot can affect the efficiency.

Either way, we do not anticipate serious problems from
nonindependence. To stress-test our results against
violations of independence, we expressly use a data-dependent pilot
for all of our experiments: namely, a weighted case-control sample
with sample points allowed to be recycled for the second-stage fit.

\subsection{Variance for a larger sample}\label{subsecAsyLargerSample}

In Section~\ref{subsecLargerSample}, we proposed increasing the size of
the local case-control subsample by multiplying all the acceptance
probabilities $a(x,y)$ by a constant $c>1$ and assigning weight
$w=ca(x,y)$ when $ca(x,y)>1$. We analyze the asymptotic
variance here as a
function of $c$. To simplify matters, suppose the model
is correctly specified and $\lambda$ is fixed at $\theta_0$.

The weighted log-likelihood for the subsample and its derivatives are then
%
\begin{eqnarray}
\ell_w(\theta) &=& \sum_{i=1}^n
z_i w_i \bigl(y_i \theta'x_i
- \log\bigl(1+e^{\theta'x_i}\bigr)\bigr),
\\
\nabla_\theta\ell_w(\theta) &=& \sum
_{i=1}^n z_i w_i
\bigl(y_i - p_\theta(x_i)\bigr)x_i,
\\
\nabla_\theta^2 \ell_w(\theta) &=& \sum
_{i=1}^n z_i w_i
p_\theta(x_i) \bigl(1-p_\theta(x_i)
\bigr)x_ix_i'.
\end{eqnarray}
Conditionally on $x$, there is a
$p(x)\cdot(c(1-p(x))\wedge1)$ chance $y=z=1$
and $w=c(1-p(x))\vee1$, where $p(x)=p_{\theta_0}(x)$. Similarly,
there is a $(1-p(x))\cdot(cp(x)\wedge1)$ chance $y=0,
z=1$, and $w= cp(x)\vee1$. We immediately obtain
%
\begin{eqnarray}
\mathbb{E}(yzw\gv x)&=&cp(1-p),\nonumber
\\
\mathbb{E}(zw\gv x) &=& 2cp(1-p),
\\
\mathbb{E}\bigl(zw^2\gv x\bigr) &\leq& c(c+1)p(1-p).\nonumber
\end{eqnarray}

The expectation and variance of the score evaluated at 0 are
%
\begin{eqnarray}
\label{eqnWtdScoreZero} \mathbb{E}\nabla_\theta\ell_w(0) &=& n \int{
\mathbb{E}\bigl(zw(y-1/2)\gv x\bigr) x \,d\mathbb{P}(x)} = 0,
\\
J = \Var\bigl(\nabla_\theta\ell_w(0) \bigr) &=& n\int{
\mathbb{E}\bigl(z^2w^2(y-1/2)^2\gv x\bigr)
xx' \,d\mathbb{P}(x)}
\\
\label{eqnWtdVar} &=& \frac{n}{4}\int{\mathbb{E}\bigl(zw^2\gv x
\bigr) xx' \,d\mathbb{P}(x)} \preceq\frac{c(c+1)}{4}
\Sigma_{\mathrm{full}}
\end{eqnarray}
and the expected Hessian is
%
\begin{equation}
H = \mathbb{E}\nabla_\theta^2\ell_w(0) =
\frac{n}{4}\int{\mathbb{E}(zw\gv x) xx' \,d\mathbb{P}(x)} =
\frac{c}{2} \Sigma_{\mathrm{full}}^{-1}.
\end{equation}
We have derived
%
\begin{equation}
H^{-1}J H^{-1} \preceq\biggl(1+\frac{1}{c} \biggr)
\Sigma_{\mathrm{full}}.
\end{equation}
For $c=1$, we recover the factor of two from (\ref{eqnAsyVar2}), but,
for example, $c=5$ we only pay 20\% increased variance relative to the
full sample.

\section{Simulations}\label{secSimulations}

Here, we compare our method to standard weighted and unweighted
case-control sampling for two-class
Gaussian models like the one considered in
Section~\ref{subsecCCBias}. The standard case-control estimates use a
50--50 split between the two classes.

\subsection{Simulation 1: Two-class Gaussian, different variances}\label{subsecSim1}

We begin with a five-dimensional two-class Gaussian simulation where the
classes have different covariance matrices.
If $X\gv Y=y \sim N(\mu_y,\Sigma_y)$,
then
%
\begin{eqnarray}\label{eqnQDA}
\log\frac{\mathbb{P}(x\gv Y=1)}{\mathbb{P}(x\gv Y=0)} &=&
-\frac{1}{2}(x-
\mu_1)'\Sigma_1^{-1}(x-
\mu_1)\nonumber\\[-8pt]\\[-8pt]
&&{} + \frac{1}{2}(x-\mu_0)'\Sigma_0^{-1}(x-\mu_0) + \operatorname{const.}\nonumber
\end{eqnarray}

Equation~(\ref{eqnQDA}) is linear if $\Sigma_1=\Sigma_0$, and quadratic
otherwise, so if the two covariance
matrices were the same the linear logistic model would be correctly
specified. In this case the model is incorrectly specified, letting
us compare the behavior of the different methods under
model misspecification.

Take $\mathbb{P}(Y=1)=1\%$, $\mu_0=0$, and $\mu_1=(1,1,1,1,4)'$. The
covariance matrices
are $\Sigma_0=\diag(1,1,1,1,9)$ and
$\Sigma_1=I_5$. Hence $f(x)$ is additive, but with a
nonzero quadratic term in $x_5$.

For our simulation, we first generate a large ($n=10^6$) sample from
the population described above. Second, we obtain a pilot model
using the weighted case-control method on $n_s=1000$ data points.
Next, we take a local case-control \mbox{sample} of size 1000
using that pilot model.

%
\begin{table}[b]
\tablewidth=300pt
\caption{Estimated bias and variance of $\hat\beta$ for each sampling method. For
$\hat\beta\in\mathbb{R}^p$, we define $\operatorname{Bias}^2 = \|
\mathbb{E}\hat\beta- \beta\|^2$ and $\Var= \sum_{j=1}^p \Var(\hat\beta_j)$}\vspace*{6pt}\label{tabMSE}
\begin{tabular*}{\tablewidth}{@{\extracolsep{\fill}}@{}l c d{1.4}d{2.6} c d{1.3}d{2.6}@{$\!$}}
\multicolumn{7}{@{}c@{}}{\textit{Simulation} 1 ($\Sigma_0\neq\Sigma_1 \Rightarrow{}$\textit{model misspecified})}\\
\hline
& & \multicolumn{1}{c}{$\bolds{\widehat{\operatorname{Bias}}{ }^2}$} & \multicolumn{1}{c}{\textbf{(s.e.)}}
& & \multicolumn{1}{c}{$\bolds{\widehat{\operatorname {Var}}}$} & \multicolumn{1}{c@{}}{\textbf{(s.e.)}}\\
\hline
LCC && 0.0049 &(0.00031) && 0.025 &(0.00059)\\
WCC && 0.023 &(0.0022) && 0.16 &(0.0038)\\
CC  && 0.15 &(0.0016) && 0.043 &(0.00096)\\
\hline\noalign{\vspace*{9pt}}
\multicolumn{7}{@{}c}{\textit{Simulation} 2 ($\Sigma_0= \Sigma_1 \Rightarrow{}$\textit{model correct})}\\
\hline
& & \multicolumn{1}{c}{$\bolds{\widehat{\operatorname{Bias}}{ }^2}$} & \multicolumn{1}{c}{\textbf{(s.e.)}}
& & \multicolumn{1}{c}{$\bolds{\widehat{\operatorname {Var}}}$} & \multicolumn{1}{c@{}}{\textbf{(s.e.)}}\\
\hline
LCC && 0.0037 &(0.0083) && 0.039 &(0.00045)\\
WCC && 0.59 &(0.064) && 1.7 &(0.017)\\
CC  && 0.06 &(0.042) && 0.87 &(0.0086)\\
\hline
\end{tabular*}\vspace*{-3pt}
\end{table}

For comparison, we obtain standard case-control
(CC) and weighted case-control (WCC) estimates. For the comparison
estimators we do not
use a sample of size 1000 again but rather use the total number of
observations seen by the LCC model or the pilot model, roughly
2000, so the LCC estimate must pay for its pilot
sample. We repeat this entire procedure 1000 times.

Table~\ref{tabMSE} shows the squared bias and variance of
$\hat\beta$ over the 1000 realizations for each of the three methods.
As expected, we face a
bias-variance tradeoff in choosing between the WCC and CC methods,
whereas the LCC method improves substantially on the bias of CC and
the variance of WCC. Standard errors for both bias and variance are
computed via bootstrapping the 1000 realizations.

More surprising is the fact that LCC enjoys smaller bias than
WCC and smaller variance than CC, dominating the other two methods on
both measures. The improvement in variance over the CC estimate is
likely
due to the conditional imbalance present in the sample, while the
improvement in bias over the WCC estimate may come from the fact that
the methods are only unbiased asymptotically and the LCC estimate is
closer to its asymptotic limiting behavior.

\subsection{Simulation 2: Two-class Gaussian, same variance}\label{subsecSim2}

Next, we simulate a two-class Gaussian model with each class having the
same variance, so that the true log-odds
function $f$ is linear. We also increase the dimension to 50
for this simulation.

Since the model is now correctly specified, all three methods are
asymptotically unbiased. However, in this case we introduce more
substantial conditional imbalance, to demonstrate the variance-reduction
advantages of local case-control sampling in that setting.

For this example, $\mathbb{P}(Y=1)=10\%$,
$\mu_1={1_{25}\choose0_{25}}$,
$\mu_0=0_{50}$, and $\Sigma_0=\Sigma_1=I_{50}$. We repeat the
procedure from Section~\ref{subsecSim1}, now with
$n_s=10^4$. Instead of generating a full sample, the full data set is
implicit and we sample directly from $\mathbb{P}_S$.

In this example, the difference between the methods is more dramatic.
Table~\ref{tabMSE} shows the squared bias and variance of the
three methods. Here, local case-control enjoys substantially better
bias than the other two methods, improving on CC more than twenty-fold.
For the correct pilot model, $\bar a(\theta_0)$ is roughly 0.005, so
the local case-control subsample size is around $n/200$. Since the model
is correctly specified, the variance is roughly twice that of logistic
regression on the full sample of size $n$. In other words,
local case-control subsampling is roughly 100 times more efficient
than uniform subsampling.

Asymptotically, all three methods are unbiased but it appears that LCC
again enjoys a smaller bias in finite samples.

\section{Web spam data set}\label{secSpam}

Relative to standard case-control sampling, local case-control
sampling is especially well-suited for data sets with significant
conditional imbalance, that is, data sets in which $y_i$ is
easy to predict for most $x_i$.

One such application is spam filtering. To demonstrate the advantages
of local case-control sampling and compare asymptotic predictions to
actual performance, we test our method on the Web Spam data available
on the LIBSVM website\footnote{\url{http://www.csie.ntu.edu.tw/\textasciitilde cjlin/libsvmtools/datasets/}.}
and originally from \citet{webb2006introducing}.
The data set contains 350,000 web pages, of which about
60\% are labeled as ``web spam,'' that is, web pages designed to manipulate
search engines rather than display legitimate content. This data set
is marginally balanced, though as we will see the conditional
imbalance is considerable.

As features, we use frequency of the 99 unigrams that appeared in at least
200 documents, log-transformed with an offset so as to reduce skew in
the features.
In this data set, the downsampling ratio $\bar a$ is around 10\%, that
is, when using a good pilot we will retain about 10\% of the observations.

Since we only have a single data set, we use subsampling as a method
to assess the sampling distribution of our estimators. In each
of 100 replications, we begin by taking a
uniform subsample of size $n={}$100,000 from the population of 350,000
documents. After obtaining 100 data sets of size $n={}$100,000,
we use the same procedure as we used in our two simulations
with $n_S={}$10,000.


Our asymptotic theory predicts that the variance of the local
case-control sampling estimate of $\theta$ should be a little more
than twice the variance using the full sample (more because the model
is misspecified and our pilot has some variance).
Because the full sample is close to marginally balanced, the standard
case-control sampling methods should do about as well as a uniform
subsample of size 20,000---that is, they should have variance roughly
5 times that of the full sample.

Note that 20,000 is roughly twice the
size of the local case-control sample, since we are counting the
pilot sample against the local case-control method. If we had a
readily available pilot model, as we would in many applications,
it would be more relevant to give the CC and WCC
methods access to
only 10,000 data points, doubling their variance relative to
the observed variance in this experiment.

\begin{figure}

\includegraphics{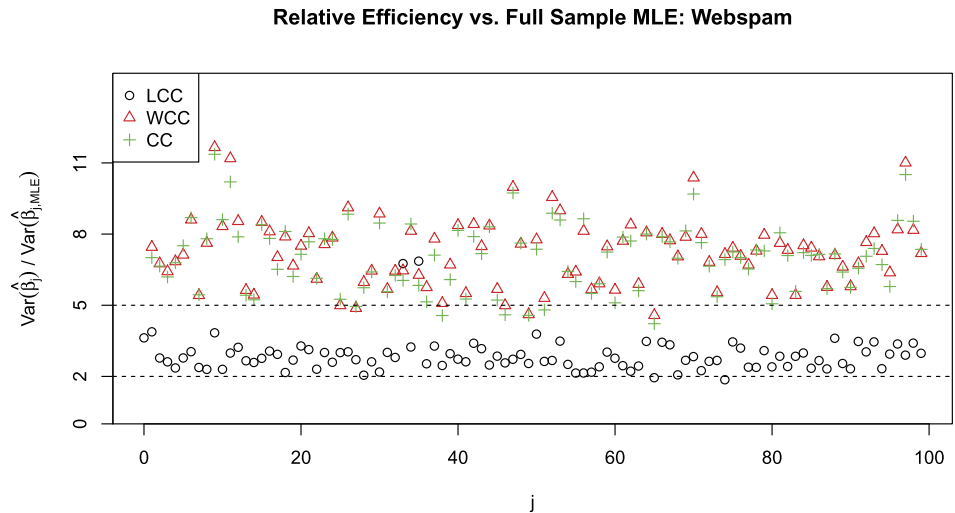}

\caption{Relative variance of coefficients for different subsampling
methods. The theoretical predictions ($2\times$ variance for local
case-control, $5\times$ variance for standard) are reasonably close
to the mark, though a bit optimistic.}
\label{figWebspamCoefs}
\end{figure}

The theoretical predictions come reasonably close in this experiment,
as shown in Figure~\ref{figWebspamCoefs}. The horizontal axis indexes
each of the 100 coefficients to be fit (there are 99 covariates and an
intercept), and the vertical axis gives the variance of each estimated
coefficient, relative to the variance of the same coefficient in a
model fitted to the full sample.

%

The magnitude of our improvement over standard case-control sampling
is substantial here, but could be much larger in a data set with
an even stronger signal. The key point is that standard case-control
methods have no way to exploit conditional imbalance, so the more
there is, the more local case-control dominates the other methods.

\section{Discussion}\label{secDiscussion}

We have shown that
in imbalanced logistic regression, we can speed up computation by
subsampling the data in a biased fashion and making a post-hoc
correction to the coefficients estimated in the subsample.
Standard case-control sampling is one such scheme, but it has two main
flaws: it has no way to exploit conditional imbalance, and
when the model is misspecified it is inconsistent for the
population risk minimizer.

Local case-control sampling generalizes standard case-control sampling
to address both flaws, subsampling with a bias
that is allowed to depend on both $x$ and $y$. When the pilot is
consistent, our estimate is consistent even under misspecification,
and if the model is correct then
local case-control sampling has exactly twice the asymptotic
variance of logistic
regression on the full data set. Our simulations suggest that local
case-control performs favorably in practice.

\subsection{Translating computational gains to statistical gains}\label{subsecAdvantages}

In the \hyperref[sec1]{Introduc-} \hyperref[sec1]{tion}, we motivated our inquiry by identifying four ways
that computational gains can translate to statistical
ones. Specifically, we suggested that computational savings can:
\begin{longlist}[(2)]
\item[(1)] enable us to experiment with and prototype
a variety of models, instead of trying only one or two,
\item[(2)] allow us to refit our models more often to adapt to changing
conditions,
\item[(3)] allow for cross-validation, bagging, boosting,
bootstrapping, or other computationally intensive statistical
procedures or
\item[(4)] open the door to using more sophisticated statistical
techniques on a compressed data set.
\end{longlist}
It is relatively clear how our proposed method can help with points (1)
and (2). As for point (3), faster fitting procedures can directly speed
up straightforward
resampling techniques like bootstrapping or cross-validation, possibly
making them feasible at scales where they previously were not. We
discuss in
Section~\ref{subsecExtens} how it can also help with boosting.

The basic method as we have described it above does not deliver on
point (4), because the pilot model and second-stage model are the same.
However, an extension of our method can help, which we discuss below.

There is no
reason in principle why the pilot model must be linear, or
belong to the same model class as the model we fit to the local
case-control sample. We can use any pilot predictions
$\tilde p(x) = \frac{e^{\tilde f(x)}}{1+e^{\tilde f(x)}}$ in the
sampling algorithm, and then
model the log-odds in the subsample quite flexibly---by a GAM,
kernel logistic regression, random forests or any other method---so
long as we can use\vspace*{1pt} offsets $-\tilde f(x_i)$ in the second-stage
procedure. For example, we could use as our pilot fit a simple model
with a few important variables explaining most of the
response, and in the second stage estimate more
complex models refining the first.

Formally, our theoretical results may not cover this use. Suppose the
second-stage model can be written as a logistic regression after
some basis expansion.\vadjust{\goodbreak} Then
consistency of the second-stage estimate requires either that the
pilot be consistent (the new variables contribute nothing to the
population fit) or that the second-stage model be correctly
specified. If neither of these assumptions holds approximately, then
our estimate could be biased---though perhaps not as biased
as case-control sampling, which is a special case of local
case-control with an intercept-only pilot.

If we are prototyping, guarantees of consistency may not be a high
priority. If they are, then as with case-control
sampling, we can repair the inconsistency of the local case-control
estimate by using a Horvitz--Thompson estimator with weights
$a_{\tilde\theta}(x_i,y_i)^{-1}$. This may come at a cost
of some added variance. It would be interesting to examine the bias
of local case-control and the variance of weighted local case-control
in this more general problem setting.

\subsection{Extensions}\label{subsecExtens}

This work suggests extensions in several directions, described below.

\subsubsection*{Indifference point other than 50\%}
In some applications
(e.g., diagnostic medical screening), a false negative may be more
costly than a false positive, or \mbox{vice-versa}. One of the
implications of the discussion in Section~\ref{subsecCCBias} is that
the Bernoulli log-likelihood implicitly places most
emphasis on approximating the log-odds well near the 0 (50\%
probability) level
curve, which may not be appropriate if the decision boundary
relevant to our application is at 10\%. In general, we would expect
to obtain a better model in the large-$n$ limit if we target the
decision boundary we care most about.

In a sense, the reason that standard case-control sampling performed
so badly in Example~\ref{ex2} of Section~\ref{subsecCCBias} is that it
targeted a level curve of ${\mathbb{P}(Y=1\gv X=x)}$ other than 50\%.
Specifically, it targeted the level curve corresponding to 50\% in the
subsampling population for equal-sampled case-control sampling, which
corresponds to the marginal $\mathbb{P}(Y=1)$ level curve in the original
population.

What happened by accident in Example~\ref{ex2} need not always be one, and it
would be interesting to generalize our procedure so as to target any chosen
decision threshold. More generally still, our indifference point
could depend on our features $x$---in online advertising, for
instance, some advertisers may be willing to pay more per click
than others.

\subsubsection*{Boosting}

In Section~\ref{subsecAdvantages} we suggested using offsets to obtain
a complex second-stage fit. Alternatively, we can obtain any fitted
log-odds function $f_s(x)$ for the sample and simply add it to the pilot
$\tilde f(x)$ to obtain an estimate for $f(x)$.

This observation suggests the possibility of iteratively fitting a
``base model'' to the subsample, then adding it to
$\tilde f(x)$ to obtain a new pilot for the next iteration.
Indeed, that iterative algorithm is closely related to the AdaBoost
algorithm of \citet{freund1997decision}. Even more similarly to
AdaBoost, we could weight each point
by $|y_i-\tilde p(x_i)|$ instead of sampling it with that probability.

\citet{friedman2000additive} show
that the AdaBoost algorithm can be thought of as fitting a logistic
regression\vspace*{1pt} model additive in base learners. In AdaBoost, the function
$F_M(x) = \sum_{m=1}^M f_m(x)$ simply records the number of
classifiers $f_m$ classifying $x$ as belonging to class $+1$ minus the
number classifying it as class $-1$, and Friedman et al. show that
$\frac{1}{2}F_M(x)$ can be thought of as approximating the log-odds of
$Y=+1$ given $X=x$.

The difference is that while AdaBoost weights the point
$(x_i,y_i)$ by\break  $e^{(2y_i-1) F_m(x_i)}$, the local case-control version
would use weights
%
\begin{equation}
\bigl|y_i-p_M(x_i)\bigr| = \frac{e^{y_i F_m(x_i)}}{1+e^{F_m(x_i)}} =
\frac
{e^{(2y_i-1)F_m(x_i)}}{1+e^{(2y_i-1)F_m(x_i)}}.
\end{equation}

Operationally, this alternative weighting scheme limits the
influence of ``outliers,'' that is, hard-to-classify points that can
unduly drive the AdaBoost fit.

\subsubsection*{Logistic regression with regularization}
In high-dimensional
settings,\break  lasso- or ridge-penalized logistic regressions are often
preferable to standard logistic regression, the model considered
here. One could use local case-control sampling with a
regularized version of logistic regression, but our asymptotic
results might need revisiting in such a case---especially in a
high-dimensional asymptotic regime [$p\gg n$ or $p/n\rightarrow
\gamma\in(0,\infty)$]. Since the high-dimensional setting
is important in modern statistics and machine learning, this bears further
investigation.

\subsubsection*{Other generalized linear models}
One way of viewing the method is
as a way of ``tilting'' the conditional distribution of $Y$ by a
linear function of $X$ in the natural parameter space so as to
enrich our subsample for more informative observations. We could use
similar tricks on other GLMs.

For instance,\vspace*{1pt} suppose we are given a Poisson variable with natural
parameter $\eta=\log\mathbb{E}Y$. By sampling with acceptance probability
proportional to $e^{\xi Y}$, we obtain (conditional on acceptance) a
Poisson with
natural parameter $\eta+ \xi$. Since Poisson variables with larger
means carry more information, this could yield a substantial
improvement over uniform subsampling.\looseness=1

If our data arise from a Poisson GLM with $\eta(x) \approx
\alpha+\beta'x$, we could generalize the local case-control scheme
by sampling $(x_i,y_i)$ with probability proportional to
$\exp\{(\xi_0-\alpha-\beta'x_i)y_i\}$, where the extra parameter
$\xi_0$ guarantees that we always tilt the conditional mean of $y_i$
upward. Similar generalizations may apply for multinomial logit and
survival models.

\begin{appendix}\label{appe}
\section{Proof of Lemma~\texorpdfstring{\lowercase{\protect\ref{lemma1}}}{1} (uniqueness of \texorpdfstring{$\theta^*$}{theta})}
Because $R_\lambda(\theta)$ is strictly convex, it is sufficient to
show that $R_\lambda(\theta)\rightarrow\infty$ as $\theta
\rightarrow\infty$ in any
direction.

Assume w.l.o.g. there is some neighborhood $N\subseteq\mathbb{R}^p$
for which
%
\begin{equation}
\label{eqnCond} \inf_{x\in N} \frac{\theta'x}{\|\theta\|} = \varepsilon
> 0,\qquad
\mathbb{P}(X\in N)>0\quad\mbox{and}\quad\mathbb{P}(Y = 1\gv X\in N) =
\pi_N < 1.\hspace*{-30pt}
\end{equation}
$h(\eta;\pi_N)$ is linear in its second argument, and is
increasing for sufficiently large $\eta$. Thus,
for large enough $\|\theta\|\varepsilon$, the population risk for
$\mathbb{P}$ is
%
\begin{eqnarray}
R(\theta) &=& \int h\bigl(\theta'x; p(x)\bigr) \,d\mathbb{P}(x)
\\
&\geq&\int_N h\bigl(\|\theta\|\varepsilon; p(x)\bigr) \,d
\mathbb{P}(x)
\\
&=& h\bigl(\|\theta\|\varepsilon; \pi_N\bigr)\mathbb{P}(X\in N) \rightarrow
\infty.
\end{eqnarray}

$\mathbb{P}_\lambda\gg\mathbb{P}$ for any $\lambda$, so (\ref{eqnCond})
holds for $\mathbb{P}_\lambda$ with the same $N$ (but a different
$\pi_N<1$). Thus,
we can repeat the same argument with $\mathbb{P}$ replaced by $\mathbb
{P}_\lambda$.

\section{Proof of Proposition~\texorpdfstring{\lowercase{\protect\ref{prop_pointwise}}}{3}
(pointwise convergence)}
Fix $\theta$ and begin by writing
%
\begin{equation}
\ell_i^\lambda= y_i(\theta-
\lambda)'x_i - \log\bigl(1+e^{(\theta
-\lambda)'x_i} \bigr).
\end{equation}

Let $z_i^\lambda$ be the Bernoulli selection decisions, generated by
comparing mutually independent $u_i\sim U(0,1)$ to the threshold
$a_\lambda(x_i,y_i)$. The $z_i^\lambda$ are independent
conditional on $\lambda$ and the data. Also, write $q_i^\lambda=
z_i^\lambda\ell_i^\lambda$, so that $\widehat Q_\lambda(\theta) =
\frac{-1}{n\bar a(\lambda)}\sum_{i=1}^n q_i^\lambda$.

By the Cauchy--Schwarz inequality, we have
%
\begin{equation}
\bigl|\ell_i^\lambda\bigr| \leq1+| \|\theta-\lambda\| \|
x_i\|.
\end{equation}

Now, for $\delta>0$ define $\Lambda_\delta= \{\lambda\dvtx
\|\lambda-\lambda_\infty\|<\delta\}$. For $\lambda\in\Lambda_1$,
we have
%
\begin{equation}
\bigl|q_i^\lambda\bigr|\leq m_i \triangleq1 + \bigl(\|\theta-
\lambda_\infty\|+1\bigr)\| x_i\|
\end{equation}
which is integrable by assumption. Finally let $\mathbb{E}_n$ denote an
average taken over indices $i=1,\ldots,n$, that is, $\mathbb{E}_n f =
\frac{1}{n}\sum_{i=1}^n f_i$. Then
%
\begin{equation}
\Qhat_{\lambda_n}(\theta) - Q_{\lambda_\infty}(\theta) = \bar a(
\lambda_n)^{-1}\mathbb{E}_n q^{\lambda_n} -
\bar a(\lambda_\infty)^{-1}\mathbb{E}q^{\lambda_\infty}.
\end{equation}

By continuity, $\bar a(\lambda_n)\stackrel{p}{\rightarrow
}\bar a(\lambda_\infty)>0$.
Therefore, it suffices to show $\mathbb{E}_n q^{\lambda_n}\stackrel
{p}{\rightarrow}\mathbb{E}q^{\lambda_\infty}$.
Because $\mathbb{E}_n q^{\lambda_\infty}\stackrel{\mathrm
{a.s.}}{\rightarrow}\mathbb{E}q^{\lambda_\infty}$ by the law of
large numbers, it
suffices equally well to show that $\mathbb{E}_n q^{\lambda_n} -
\mathbb{E}_n
q^{\lambda_\infty}\stackrel{p}{\rightarrow}0$.

Now fix $\varepsilon>0$ and take $K$ large enough that $\mathbb
{E}(m\mathbf{1}_{m>K})<\varepsilon$.
For $\lambda_n\in\Lambda_1$ we have
%
\begin{equation}
\bigl\llvert\mathbb{E}_n q^{\lambda_n}- \mathbb{E}_n
q^{\lambda_\infty
}\bigr\rrvert\leq\bigl\llvert\mathbb{E}_n
\bigl(q^{\lambda_n}-q^{\lambda_\infty}\bigr)\mathbf{1}_{m\leq K}\bigr
\rrvert+ 2\mathbb{E}_nm\mathbf{1}_{m>K}.
\end{equation}

With probability one the second term is eventually less than
$2\varepsilon$.
Further, for
$\lambda_n\in\Lambda_\delta$, we have
%
\begin{eqnarray}
\bigl|q_i^{\lambda_n}-q_i^{\lambda_\infty}\bigr| &=&
\tfrac{1}{2}\bigl\llvert\bigl(z_i^{\lambda_n}-z_i^{\lambda_\infty}
\bigr) \bigl(\ell_i^{\lambda_n}+\ell_i^{\lambda_\infty}
\bigr) + \bigl(z_i^{\lambda_n}+z_i^{\lambda_\infty}
\bigr) \bigl(\ell_i^{\lambda_n}-\ell_i^{\lambda_\infty}
\bigr) \bigr\rrvert
\\
&\leq& \bigl|z_i^{\lambda_n}-z_i^{\lambda_\infty}\bigr|m_i
+ \delta\|x_i\|.
\end{eqnarray}

Now, write
%
\begin{equation}
d_i = \bigl|z_i^{\lambda_n}-z_i^{\lambda_\infty}\bigr|m_i
\mathbf{1}_{m_i\leq K}.
\end{equation}

$z_i^{\lambda_n}\neq z_i^{\lambda_\infty}$ iff $u_i$ lies
between $a_{\lambda_n}(x_i,y_i)$ and $a_{\lambda_\infty}(x_i,y_i)$.
Hence, conditionally on $\lambda_n$ and the data, the $d_i$ are mutually
independent nonnegative random variables bounded by $K$ with means
%
\begin{equation}
\mu_i=\bigl|a_{\lambda_n}(x_i,y_i)-a_{\lambda_\infty
}(x_i,y_i)\bigr|m_i
\mathbf{1}_{m_i\leq
K} < \delta K^2
\end{equation}
since $\nabla_\lambda a_\lambda(x_i,y_i) \leq\|x_i\| < m_i$.

Continuing, we have
%
\begin{eqnarray}
\bigl\llvert\mathbb{E}_n \bigl(q^{\lambda_n}-q^{\lambda_\infty}
\bigr)\mathbf{1}_{m\leq K}\bigr\rrvert&\leq&\mathbb{E}_n(d-\mu)
+ \mathbb{E}_n\mu+ \delta\mathbb{E}_n\| x\|
\mathbf{1}_{m\leq K}
\\
&\leq&\mathbb{E}_n(d-\mu) + \delta K^2 + \delta K.
\end{eqnarray}

Conditioning on $\lambda$ and $\{(x_i,y_i)\}$,
the first term is a sum of independent zero-mean random variables
that are bounded in absolute value by $K$. By Hoeffding's inequality,
%
\begin{equation}
\label{eqnHoeffding} \mathbb{P} \Biggl( \Biggl\llvert\frac{1}{n}\sum
_{i=1}^n{d_i-\mu_i}\Biggr\rrvert\geq\varepsilon\Big|\lambda_n,
\bigl\{(x_i,y_i)\bigr\} \Biggr) \leq2\exp\bigl[-n
\varepsilon^2/\bigl(2K^2\bigr) \bigr].
\end{equation}

Since this bound is deterministic, the same applies to the
unconditional probability that $\mathbb{E}_n(d-\mu)$ is large. Take
$\delta= \varepsilon/(K+K^2)$.
With probability tending to~1, $\lambda_n\in\Lambda_\delta$ and the
event in (\ref{eqnHoeffding}) holds, in which case
%
\begin{equation}
\bigl|\mathbb{E}_n \bigl(q^{\lambda_n}- q^{\lambda_\infty}\bigr)\bigr| \leq4
\varepsilon.
\end{equation}

Since $\varepsilon$ was arbitrary, the proof is complete.

\section{Proof of Theorem~\texorpdfstring{\lowercase{\protect\ref{thm_condl_var}}}{6}
[distribution~of~\texorpdfstring{$\hat\theta-\bar\theta(\lambda)$}{theta - theta(lambda)}]}
By the mean value theorem, we have for each $n$
%
\begin{equation}
\nabla_\theta\widehat Q_{\lambda_n}(\hat\theta_n) =
\nabla_\theta\widehat Q_{\lambda_n}\bigl(\bar\theta(
\lambda_n)\bigr) + \nabla_\theta^2 \widehat
Q_{\lambda_n}(\phi_n) \bigl(\hat\theta_n-\bar\theta(
\lambda_n) \bigr),
\end{equation}
where $\phi_n$ is some convex combination of $\hat\theta_n$
and $\bar\theta(\lambda_n)$. Noting that the LHS is by
definition 0
and rearranging, we obtain
%
\begin{equation}
\sqrt{n} \bigl(\hat\theta_n-\bar\theta(\lambda_n) \bigr)
= \nabla_\theta^2 \widehat Q_{\lambda_n}(
\phi_n)^{-1} \cdot\sqrt{n} \nabla_\theta\widehat
Q_{\lambda_n}\bigl(\bar\theta(\lambda_n)\bigr).
\end{equation}
If we can show the first factor tends in probability to
$\nabla_\theta^2 Q_{\theta^*}(\theta^*)^{-1}$ and the second tends in
distribution to $N (0,\bar a(\theta^*)^{-1}J(\theta^*,\theta^*) )$,
then by Slutsky's
theorem we have the desired result.

Using the Skorokhod construction define a joint
probability space for $\lambda_n$ such that $\lambda_n\stackrel
{\mathrm{a.s.}}{\rightarrow}
\theta^*$. We will condition on the sequence $\lambda_n$ and use a
triangular array central limit theorem for the random variables
%
\begin{eqnarray}
g_{ni} &=& \frac{z_{ni}}{\bar a(\lambda_n)} \biggl(y_{i}-
\frac{e^{(\bar\theta(\lambda_n) - \lambda
_n)'x_i}}{1+e^{(\bar\theta(\lambda_n) - \lambda_n)'x_i}} \biggr)x_i
\\
&=& \frac{z_{ni}}{\bar a(\lambda_n)}\nabla_\theta\ell(\theta-\lambda
_n;x_{i},y_{i})
\Big|_{\theta=\bar\theta(\lambda_n)}.
\end{eqnarray}
Because $\lambda_n$ is independent of the data,
$\mathbb{E}(f(g_{ni})\gv\lambda_n,z_{ni}=1) = \mathbb{E}_{\lambda
_n}(f(g_{ni}))$
for any $f$. The triangular array CLT applies since
%
\begin{eqnarray}
\qquad \mathbb{E}(g_{ni}\gv\lambda_n) &=& 0,
\\
\Var(g_{ni}\gv\lambda_n) &=& \mathbb{E} \bigl[
\Var(g_{ni}\gv\lambda_n,z_{ni})\gv\lambda
_n \bigr]
\\
&=& \mathbb{P}(z_{ni}=1\gv\lambda_n)\bar a(
\lambda_n)^{-2} \Var_{\lambda_n}\bigl(
\nabla_\theta\ell\bigl(\bar\theta(\lambda_n)-
\lambda_n;x_{ni},y_{ni}\bigr)\bigr)
\\
&=& \bar a(\lambda_n)^{-1} J\bigl(\bar\theta(
\lambda_n),\lambda_n\bigr)
\\
&\stackrel{\mathrm{a.s.}} {\rightarrow}& \bar a\bigl(\theta^*
\bigr)^{-1}J\bigl(\theta^*,\theta^*\bigr).
\end{eqnarray}
Therefore, defining $S_n = n^{-1/2}\sum_{i=1}^n g_{ni}$ and
$Z=N(0,a(\theta^*)^{-1}J(\theta^*,\theta^*))$, the CLT tells us
$\mathbb{P}(S_n\in A\gv\lambda_n)\rightarrow\mathbb{P}(Z\in A)$ whenever
$\lambda_n\rightarrow\theta^*$, which we assumed occurs with probability
1. By dominated convergence, we also have ${\mathbb{P}(S_n\in
A)\rightarrow\mathbb{P}(Z\in A)}$.

Next we turn to the Hessian. We have
$\hat\theta_n\stackrel{p}{\rightarrow}\theta^*$ by Theorem~\ref{thmConsistency}, so
$\phi_n\stackrel{p}{\rightarrow}\theta^*$ as well. Writing
%
\begin{equation}
h_i^{\theta,\lambda} = \frac{e^{(\theta-\lambda)'x_i}}{(1+e^{(\theta
-\lambda)'x_i})^2} x_ix_i'
z_i^{\lambda}
\end{equation}
we need to show that
%
\begin{equation}
\bar a(\lambda_n)^{-1} \bigl(\mathbb{E}_n
h^{\phi_n,\lambda_n} \bigr)^{-1} \stackrel{p} {\rightarrow}\bar a\bigl(
\theta^*\bigr)^{-1} \bigl(\mathbb{E}h^{\theta^*,\theta^*}
\bigr)^{-1}.
\end{equation}
Note that $\|h_i^{\theta,\lambda}\|_F\leq\|x_i\|^2$, which is
integrable; hence $\mathbb{E}_n h^{\theta^*,\theta^*} \stackrel
{p}{\rightarrow}\mathbb{E}
h^{\theta^*,\theta^*} = H(\theta^*,\theta^*)\succ0$.
Since
$\bar a$ is continuous and strictly positive, and $\lambda
_n\stackrel{p}{\rightarrow}
\theta^*$, it suffices to show that
%
\begin{equation}
\bigl\llVert\mathbb{E}_n h^{\phi_n,\lambda_n} - \mathbb{E}_n
h^{\theta^*,\theta^*}\bigr\rrVert_F \stackrel{p} {\rightarrow}0.
\end{equation}
Note that $h_i^{\theta^*,\theta^*}=\frac{1}{4}x_ix_i'$, and define
${w_{ni}=\frac{e^{(\phi_n-\lambda_n)'x_i}}{(1+e^{(\phi_n-\lambda
_n)'x_i})^2}}$.\vadjust{\goodbreak}

Following the structure of the proof of Proposition~\ref{prop_pointwise}, take $K$ large enough that
$\mathbb{E}\|x\|^2\mathbf{1}_{\|x\|> K} < \varepsilon$ and truncate
the $h_i$:
%
\begin{eqnarray}
\nonumber && \bigl\llVert\mathbb{E}_n h^{\phi_n,\lambda_n} -
\mathbb{E}_n h^{\theta
^*,\theta^*}\bigr\rrVert_F
\\
&&\qquad \leq \bigl
\llVert\mathbb{E}_n \bigl(h^{\phi_n,\lambda_n} - h^{\theta^*,\theta^*}
\bigr) \mathbf{1}_{\|x\|\leq K}\bigr\rrVert_F
\\
\nonumber &&\quad\qquad{} + \bigl\llVert\mathbb{E}_n \bigl(h^{\phi_n,\lambda_n} -
h^{\theta^*,\theta^*} \bigr)\mathbf{1}_{\|x\|> K}\bigr\rrVert_F
\\
&&\qquad \leq K^2\mathbb{E}_n\bigl\llvert w_{n}z_n^{\lambda_n}-
\tfrac{1}{4}z_n^{\theta^*} \bigr\rrvert
\mathbf{1}_{\|x\|\leq K} + 2\mathbb{E}_n \|x\|^2
\mathbf{1}_{\|x\|>K}.
\end{eqnarray}
The second term is eventually less than $2\varepsilon$. Now,
$w_{ni}-\frac{1}{4}$ is small, because
%
\begin{equation}
\biggl\llvert\frac{d}{d\eta} \biggl(\frac{e^\eta}{(1+e^\eta)^2} \biggr
)\biggr\rrvert
= \biggl\llvert\frac{e^\eta(e^\eta-1)}{(1+e^\eta)^3}\biggr\rrvert\leq
\frac{e^\eta}{(1+e^\eta)^2} \leq
\frac{1}{4}.
\end{equation}
Hence, by Cauchy--Schwarz
%
\begin{equation}
\bigl\llvert w_{ni}-\tfrac{1}{4}\bigr\rrvert\leq
\tfrac{1}{4}\|\phi_n-\lambda_n\| \|x_i
\|.
\end{equation}
So on the event $\{\max
\|\lambda_n-\theta^*\|,\|\phi_n-\theta^*\| < \delta\}$, we have
%
\begin{eqnarray}
&&\mathbb{E}_n\bigl\llvert w_{n}z_n^{\lambda_n}-
\tfrac{1}{4}z_n^{\theta^*} \bigr\rrvert
\mathbf{1}_{\|x\|\leq K}
\\
&&\qquad  = \tfrac{1}{2}\mathbb{E}_n\bigl\llvert
\bigl(z^{\lambda_n}-z^{\theta^*}\bigr) \bigl(w_n+
\tfrac{1}{4} \bigr) + \bigl(z^{\lambda_n}+z^{\theta^*}\bigr)
\bigl(w_n-\tfrac{1}{4} \bigr)\bigr\rrvert
\mathbf{1}_{\|x\|\leq K}
\\
&&\qquad  \leq\mathbb{E}_n\bigl|z^{\lambda_n}-z^{\theta^*}\bigr|\mathbf
{1}_{\|x\|\leq K} + \delta K.
\end{eqnarray}
Finally, we can bound the first term exactly as we did in the proof
of Proposition~\ref{prop_pointwise}, defining $d_i =
|z_i^{\lambda_n}-z_i^{\theta^*}|K^2\mathbf{1}_{\|x_i\|\leq K}$ and
$\mu_i=\mathbb{E}(d_i
\gv x_i,y_i,\lambda_n) \leq\delta K^3$. The same argument implies
$\mathbb{P}(\mathbb{E}_n(d-\mu)\geq\varepsilon) \leq2\exp
[-n\varepsilon^2/(2K^4)]$, so as $n\rightarrow\infty$
we have with probability approaching 1,
%
\begin{eqnarray}
\qquad \bigl\llVert\mathbb{E}_n \bigl(h^{\phi_n,\lambda_n} - h^{\theta
^*,\theta^*}
\bigr)\bigr\rrVert_F &\leq&\mathbb{E}_n(d-\mu) +
\mathbb{E}_n\mu+ \delta K^3 + 2\mathbb{E}_n
\|x\|^2\mathbf{1}_{\|x\|>K}
\\
&\leq& 3\varepsilon+ 2\delta K^3
\end{eqnarray}
so taking $\delta<\varepsilon/K^3$, the right-hand side is less than
$5\varepsilon$.
\end{appendix}

\section*{Acknowledgements}
The authors are grateful to Jerome Friedman for suggesting that we
investigate the bias of case-control sampling, and
to Nike Sun for helpful comments and suggestions.


%

\printaddresses
\end{document}